\documentclass[aip, reprint, floats, floatfix]{revtex4-1}

\usepackage{graphicx}
\usepackage{dcolumn}
\usepackage{bm}
\usepackage[table]{xcolor}
\usepackage{ctable}
\usepackage{enumitem} 
\usepackage{amsfonts} 
\usepackage{amsmath} 

\usepackage{natbib}

\begin{document}

\title[A dynamical perspective for network analysis]
{An integrative dynamical perspective for graph theory and the analysis of complex networks }

\author{Gorka Zamora-L\'opez}  \email{gorka@Zamora-Lopez.xyz}
\affiliation{Center for Brain and Cognition, Pompeu Fabra University, Barcelona, Spain.}
\affiliation{Department of Information and Communication Technologies, Pompeu Fabra University, Barcelona, Spain.} 

\author{Matthieu Gilson} 
\affiliation{Institut des Neurosciences de la Timone, CNRS-AMU, Marseille, France.}
\affiliation{Institut des Neurosciences des Systemes, INSERM-AMU, Marseille, France.}


\begin{abstract} 	
Built upon the shoulders of graph theory, the field of complex networks has become a central tool for studying real systems across various fields of research. Represented as graphs, different systems can be studied using the same analysis methods, which allows for their comparison. Here, we challenge the wide-spread idea that graph theory is a universal analysis tool, uniformly applicable to any kind of network data. Instead, we show that many classical graph metrics---including degree, clustering coefficient and geodesic distance---arise from a common hidden propagation model: the discrete cascade. 
From this perspective, graph metrics are no longer regarded as combinatorial measures of the graph, but as spatio-temporal properties of the network dynamics unfolded at different temporal scales. Once graph theory is seen as a model-based (and not a purely data-driven) analysis tool, we can freely or intentionally replace the discrete cascade by other canonical propagation models and define new network metrics. This opens the opportunity to design---explicitly and transparently---dedicated analyses for different types of real networks by choosing a propagation model that matches their individual constraints. In this way, we take stand that network topology cannot always be abstracted independently from network dynamics, but shall be jointly studied. Which is key for the interpretability of the analyses.
The model-based perspective here proposed serves to integrate into a common context both the classical graph analysis and the more recent network metrics defined in the literature which were, directly or indirectly, inspired by propagation phenomena on networks.
\end{abstract}


\maketitle

\clearpage
\newpage
\mbox{~}
\clearpage
\newpage
\section{Introduction}

Built upon the shoulders of graph theory, the field of complex networks has become a central tool for studying real systems across many different fields of research, e.g. sociology~\cite{Borgatti2009}, epidemiology~\cite{Kiss_EpidemicBook}, neuroscience~\cite{Kaiser_Review_2007,Zamora_FrontReview_2011,Baronchelli_Review_2013,Papo_GreatExpectations_2014}, biology~\cite{Jeong2000,Junker_BookNets}, chemistry~\cite{Wickramasinghe_Chimera_2013, Bick_Chimera_2017} and telecommunications~\cite{Broder2000}. 
The success of graph theory to permeate through such a diversity of domains lies on its simplified representation. In the eyes of graph theory, a system of interacting elements is reduced to nodes and edges. A graph is an abstract manner to describe empirical systems which provides them with a ``form'' that is mathematically tractable, thus allowing to uncover their hidden architecture and to quantify how this architecture is related to---or is affected by---the functions of the real system. 
Despite its immense success, the simplicity of graph analysis is at the same time its major limitation. Reducing a real system into a graph implies discarding much of the information needed to understand it. As beneficial as it is to count with a simplified representation and having a common toolbox for all networks, the final step of the analysis is to translate the outcomes of the graph metrics back into interpretations that make sense in the context of the real system. This step---from metrics to interpretation---is prone to personal creativity, all the more when the simplifications made in first place were substantial. 

Significant efforts have been devoted in the past to study the bidirectional relation between network architecture and dynamics on networks. The literature can be divided into three classes of studies. 
First, investigations that aim at explaining how the architecture of a network, or specific structural features, influence the collective dynamics happening on a network~\cite{Arenas_Review_2008,Barrat_Book,Masuda_ReviewWalks_2017,Ji_PropagationReview_2023}. For example, how the degree distribution, the clustering coefficient or the presence of motifs affect the synchrony between nodes. 
Second, studies which---in opposition---aim at revealing the unknown organization of a network by running diffusion, propagation or navigation processes on the network~\cite{Yang_Walking_2005,Arenas_SynchScales_2006,Rosvall_Infomap_2008,Boguna_Navigability_2009,Delvenne_StabilityComms_2010,Gilson_EstimationEC_2016,Arnaudon_Centrality_2020}. For example, to identify communities or to define the centrality of nodes by observing the behavior of random walkers on a network.
And third, studies of network inference in which the connectivity itself is not (completely) known but empirical recordings are available of the network dynamics, e.g., the spiking activity of neurons. Such inference methods aim at guessing the underlying structural or effective links from the observed signals~\cite{Wu_topologies_2011,Bianco_NetInference_2016,Wai_NetRecovery_2019,Asilani_Inference_2020}.

The present Perspective Article is enclosed within the second class of studies, namely, that of employing simple propagation dynamics to describe the network structure. Here, we will point out that the relation between graphs and dynamics is not only a matter of practical interest but that a foundational correspondence exists between the two. We show that graph analysis can be reformulated from the perspective of dynamical systems, by exposing that classical graph metrics, e.g., clustering coefficient and geodesic distance, arise from a simple but common propagation dynamical model: a cascade of discrete agents, which is also discrete in time and rapidly diverges. From this dynamical perspective, graph metrics are no longer regarded as combinatorial measures of the graph but as spatio-temporal properties of the network, unfolded at different temporal scales after unit perturbations are applied at the nodes. Here, by \emph{unit perturbation} we mean an external stimulus of unit amplitude and short duration whose effect propagates throughout the network.

Exposing this dynamical viewpoint is relevant for several reasons and opens new opportunities for the study of complex networks in a more pragmatic manner. 
First, it allows to conceive graph analysis as a model-based analysis toolbox instead of a data-driven one. Given that classical (combinatorial) graph metrics implicitly assume a discrete and divergent propagation as the model to describe the interactions occurring in the network, it affects and sometimes undermines the interpretations we derive from their outcomes. However, in a model-based analysis we are free to replace the discrete cascade by other propagation models and define new network metrics. This opens the possibility to calibrate network analyses by choosing a minimal---canonical---propagation model that respects the fundamental constraints of the real system of interest; thus balancing between simplicity and interpretability.
Second, the shift from a data-driven to a model-based analysis also allows us to frame into a common context both the classical graph metrics and the more recent approaches to describe complex networks that are---one way or another---inspired by dynamics. Despite the fact that those methods were introduced independently from each other~\cite{Katz_Centrality_1952,Page_PageRank_1998,Yang_Walking_2005,Estrada_Communicability_2008,Rosvall_Infomap_2008,Boguna_Navigability_2009,Delvenne_StabilityComms_2010,Gilson_DynCom_2018,Arnaudon_Centrality_2020}, here we disclose that in reality they form a family of methods. Each method is rooted on a specific canonical propagation model and therefore each method serves a different range of applications.

The paper is organized as follows. Section~II describes the dynamical formulation of graph metrics as emerging from a discrete cascade. Section~III illustrates that networks exhibit different faces depending on the propagation model employed to observe them. This highlights the need for making explicit the dynamical assumptions underlying network analyses. 
Section~IV presents novel results---using a continuous propagation model---to demonstrate the advantages of a dynamical approach to network analysis. In particular, to generalize the concept of distance and to compare across networks.
Last, Section~V revisits past literature to define network metrics based on propagation phenomena or diffusion. We clarify the similarities and the differences between those proposals, pinpointing the underlying model in each case.
The present Article focuses mainly on binary and undirected graphs in order to reliably formalize the relation between graph metrics and propagation dynamics. However, the approach proposed here naturally extends to weighted and directed networks as illustrated, e.g., by the network normalization examples in Section~IV.

\section{A dynamical representation of classical graph metrics} 	\label{sec:2}

\begin{figure*}[ht!]
	\centering
	\includegraphics[width=1.0\textwidth,clip=]{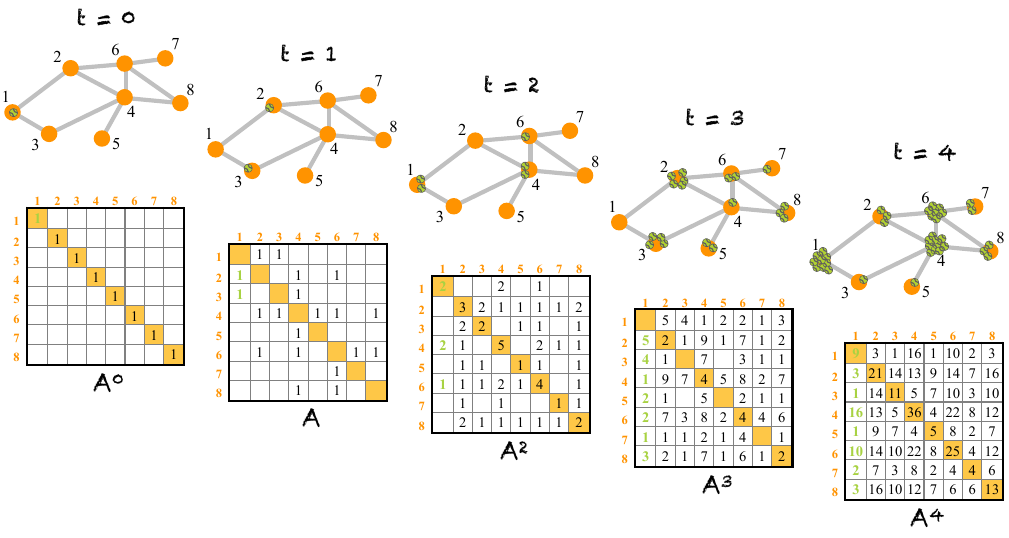}
 	\caption{		\label{fig:Figure1}
	{\bf Representation of the discrete cascade behind graph metrics.}
	From the perspective of graph theory, the powers of the adjacency matrix $A^l$ represent the number of all (non-)Hamiltonian paths of length $l$ between two nodes. From a dynamical perspective, however, matrices $A^t$ encode the number of particles found in node $i$ (row index, target node) at any given time $t$, due to one particle initially seeded at node $j$ (column index, source node) at time $t=0$; assuming the propagation of particles is governed by a discrete cascade. In the illustration, matrix entries in green color are the number of particles at target nodes $i = 1, \ldots, n$ due to initial seed at node $j=1$. Entries of value ``0'' are left blank for visualization purposes.
	} 
\end{figure*}

For graph theory, all the relevant information about a network is encoded in the adjacency matrix $A$. Hence, graph analysis consists of applying different metrics on $A$ in order to reveal the \emph{shape} of the network---its architecture. This process is similar to building a puzzle because there is no single metric which conveys all the necessary information needed to fully understand the network. Each graph metric provides a useful but incomplete piece of information and only by integrating several pieces together we can understand how the network is organized.
In this article we will follow the indexing convention of dynamical systems in which $A_{ij} = 1$ if there is a link pointing from (source) node $j$ to (target) node $i$. Hence, the columns of $A$ represent the outputs from a node and the rows its inputs. Wherever possible, $i$ will be used only as a row index and $j$ as a column index, while summations (either column or row) will involve index $k$. 

Historically, graph theory has been formalized as a branch of combinatorial mathematics. However, graph metrics are nowadays rarely evaluated using combinatorial algorithms. Instead, graph metrics are usually computed by navigating the graph via depth-first-search (DFS) or breath-first-search (BFS) algorithms and applying different rules along the process. From a dynamical standpoint, DFS and BFS represent two different propagation processes. Depth-first search corresponds to a conservative dynamical process in which a single agent explores the entire graph jumping from one node to another through the links. A similar approach widely employed in the recent literature is that of random walks. The difference is that in a DFS the agent follows a pre-established ordering that depends on the labelling of the nodes, while in random walks the agent randomly chooses which of the neighbors to visit in the following iteration.

A BFS represents a non-conservative cascading process of discrete particles. In a BFS, for every particle sitting on a node $j$ at time $t$, the process generates at $t+1$ one new particle for each of the neighbors of $j$. Therefore, BFS is non-conservative because the number of particles rapidly grows with time. 
Such a cascade is illustrated in Fig.~\ref{fig:Figure1} in a graph of $n=8$ nodes and a single particle (a tennis ball) starting from $j=1$. This has two neighbors, $i=2$ and $i=3$, who receive one ball each at time $t=1$. In the following iteration $i=1$ receives two balls---one per neighbor, $i=6$ receives one ball and $i=4$ receives two. At $t=3$ the number of balls grows from five to eighteen. Both $i=2$ and $i=3$ receive two balls from $j=1$ and two more from $j=4$. At each iteration, every node receives one new ball per each ball in its neighbors.

Without a queue to remember the nodes visited, the system that describes the cascade behind BFS in a graph is the discrete mapping $f: \mathbb{N}^n \to \mathbb{N}^n$ of the form:
\begin{equation} 	\label{eq:Cascade1}
	\mathbf{x}_{t} = A \, \mathbf{x}_{t-1} ,
\end{equation}
where $\mathbf{x}_t$ is the state (column) vector of shape $n \times 1$. The values $x_{i,t} \in \mathbb{N}$ represent the number of particles found in node $i$ at time $t$. Usually, a BFS-like process starts with a single particle at a selected node, as the example in Fig.~\ref{fig:Figure1} initiated from $j=1$. If we assume instead that the process starts with one particle per node, the initial conditions are given by the column vector of unit entries $\mathbf{x}_0^T = \mathbf{1}^T = (1,1, \ldots, 1)$. The solutions of the discrete cascade at $t > 0$ are thus obtained recursively:
\begin{eqnarray} 
	\mathbf{x}_1  & = &  A \, \mathbf{x}_0 = A\, \mathbf{1} \, , 
	\nonumber \\ 
	\mathbf{x}_2  & = &  A \, \mathbf{x}_1 = A \left( A \mathbf{x}_0) \right) = A^2 \mathbf{x}_0 = A^2 \, \mathbf{1} \, ,	
	\nonumber \\
	\mathbf{x}_3  & = &  A \, \mathbf{x}_2 = A \left(A \left( A \mathbf{x}_0 \right)\right)  =  A^3 \mathbf{x}_0 = A^3 \, \mathbf{1}	 \, ,
	\nonumber \\
			& \vdots & 	\nonumber \\
	\mathbf{x}_t  	& = &  A^t \, \mathbf{x}_0 = A^t \, \mathbf{1} \, .   
	\label{eq:Cascade2}
\end{eqnarray}
The recursive nature of the process implies that the solution at any time $t$ is trivially determined by two quantities: the initial conditions $\mathbf{x}_0$ and the powers $A^t$ of the adjacency matrix $A$ acting as the propagator (kin to the \emph{Green's function}). In this case, the values $\left( A^t \right)_{ij}$ are the number of particles found in node $i$ at time $t$, due to the single particle seeded at $j$ at $t=0$. More generally, $\left( A^t \right)_{ij}$ can be interpreted as the \emph{temporal response} of node $i$ to a unit perturbation (a stimulus of unit amplitude) applied on $j$ at time $t=0$. This conditional pair-wise response encompasses all network effects from $j$ to $i$, acting at different time scales, along all paths of different lengths.

At this point, a connection can be drawn between graph theory---as a combinatorial method---and a dynamical process that implements the calculation of graph metrics. From graph theory it is well known that the powers of the adjacency matrix, $A^l$, encode the number of Hamiltonian and non-Hamiltonian paths of length $l$ between two nodes. For example, the entry $(A^3)_{ij}$ represents the number of paths of length $l=3$ starting at node $j$ that reach node $i$. If $i=j$, then $(A^3)_{ii}$ is the number of triangles (cycles) in which $i$ participates. From a purely combinatorial point of view, counting and identifying all possible paths of a given length is a difficult problem to tackle for large networks since the number of branches rapidly grow with length $l$. 
However, from the dynamical point of view it is a rather trivial exercise. As Eq.~(\ref{eq:Cascade2}) reveals, the combinatorial problem is equivalent to iteratively calculating the propagation of a discrete cascade in the network until time horizon $t=l$. Our aim here is to show that more than a coincidence, this dynamical equivalence and interpretation is common for the most popular graph metrics, albeit often implicit or hidden.

To this end, we note that all the relevant information needed to characterize the network and to define graph metrics is unfolded---through the dynamics---from the adjacency matrix $A$ onto the response matrices $\mathcal{R} = \{ A^0, A^1, A^2, A^3, \ldots, A^t \}$. In Appendix~1 we show that node degree, matching index, clustering coefficient and geodesic distance can be derived from the matrices $\mathcal{R}_t$. From these derivations we learn three lessons. 
First, in this dynamical perspective, graph metrics are not combinatorial attributes of the graph but spatio-temporal properties of the network's response to external (unit) stimuli. 
Second, although the discrete cascade is a divergent system, graph metrics only reflect the properties of the network dynamics at short times. Both the degree and the matching index are attributes of the cascade at $t=2$, and the clustering coefficient is a network feature of time $t=3$. The geodesic distance is the only metric that may result from the cascading at longer times (up to $t = n-1$ if the graph is connected). But for real networks that exhibit small-world properties, it spans only for $t \ll n$. 
And third, graph metrics are easy to generalize for arbitrary time-scales. For example, we realize that the degree and the clustering coefficient both reflect the self-response of one node on itself at times $t=2$ and $t=3$ respectively. That is, the degree is computed from the diagonal entries $\left( A^2 \right)_{ii}$ and clustering from $\left( A^3 \right)_{ii}$. Thus, it would be easy to extend these and evaluate the self-influence---or recurrence of the inputs---at any time $t > 3$ from $\left( A^t \right)_{ii}$.

\section{Propagation model selection for personalized network analysis} 		\label{sec:3}

\begin{figure*}[ht!]
	\centering
	\includegraphics[width=1.0\textwidth,clip=]{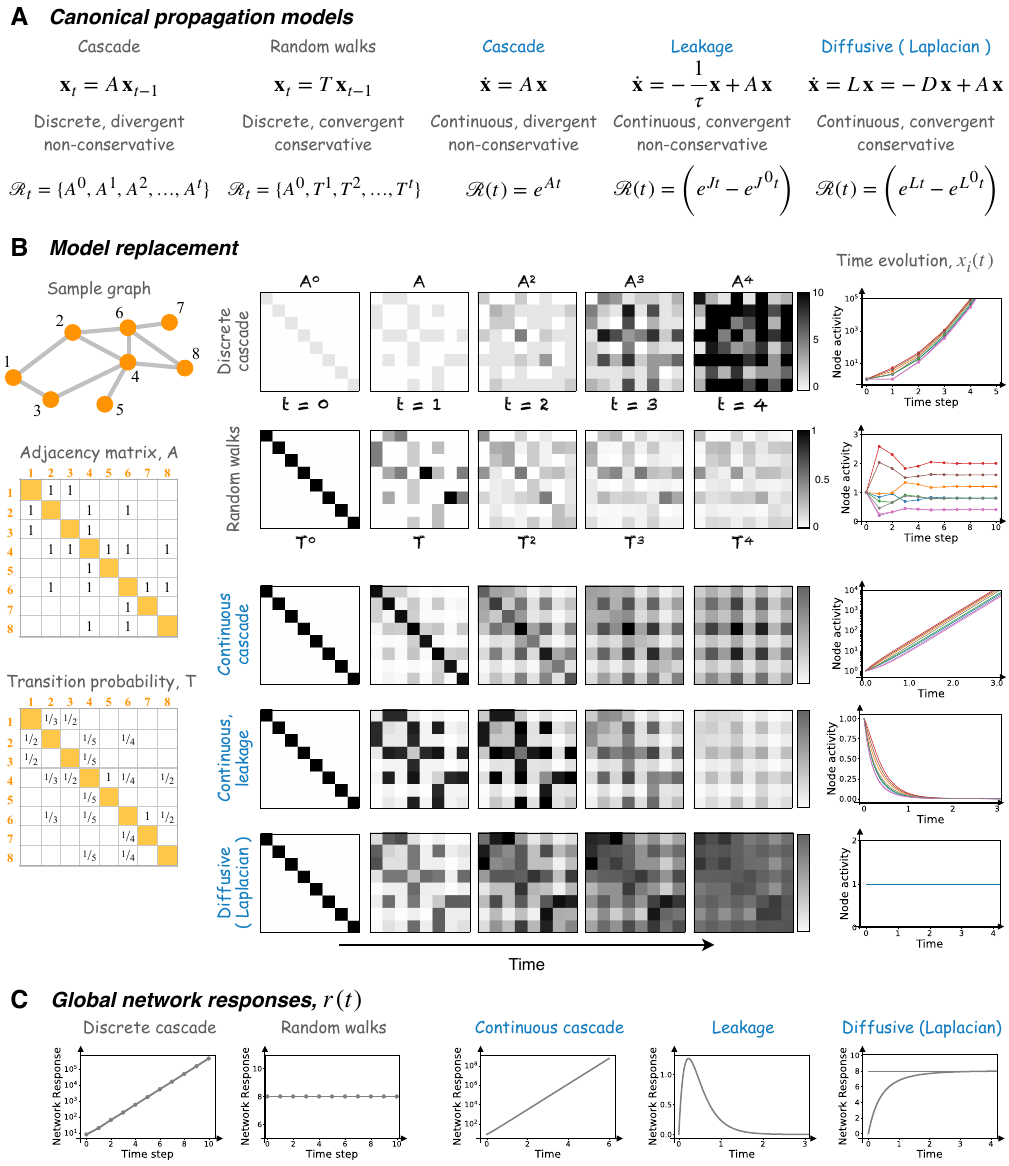}
 	\caption{		\label{fig:Figure2}
	{\bf Characterizing networks in the light of different dynamical processes.} 
	{\bf A} Equations, basic conditions and temporal response matrices of five canonical propagation models, two discrete (gray) and three continuous (blue). 
	{\bf B} Illustration of model replacement to derive network metrics applied to a sample graph of eight nodes. Each row represents the response matrices at different times for one of the five canonical models (grayscale matrices, from left to right). In all cases, the initial matrix is diagonal representing a unit perturbation at all nodes. Except for the continuous cascade, the color scale of the matrices is adjusted to the same range. Due to the explosive growth of the continuous cascade, matrices are shown for adjusted color scales in order to highlight the resulting patterns. Panels on the right hand side show the temporal solutions $x_i(t)$ for the eight nodes in the graph, highlighting the divergent, conservative or decaying nature of the models.
	{\bf C} Global network responses $r(t)$ over time, quantifying the sum of all pair-wise responses $\mathcal{R}_{ij}(t)$ at each time step $t$. 
	} 
\end{figure*}

The derivations in Sec.~\ref{sec:2} and Appendix~1 allowed us to establish a fundamental relation between graph theory and network dynamics, by showing that usual graph metrics can be derived from a common propagation model. And accordingly, they can be re-interpreted as spatio-temporal properties of the network responses to initial unit stimuli. Although defining the graph metrics from the set of response matrices $\mathcal{R} = \{ A^0, A^1, A^2, A^3, \ldots, A^t \}$ may seem a complication, this dynamical perspective brings two important implications. 
On the one hand, it reveals that when applying classical graph metrics, we are assuming ``as if'' the discrete cascade were the appropriate dynamical model to describe the real network under study. Given the number of empirical systems studied as networks, it is unrealistic to assume that one propagation model serves to characterize and interpret all of them. On the other hand, it opens the door to alleviate this problem by developing a family of graph analysis methods of different \emph{flavors}. Once recognized as a model-based analysis tool, we are free to replace the underlying propagation model and design analyses that are better suited for individual real networks, or specific domains of real networks. 

We envision that in the future, before performing a network study, users will first identify the fundamental constraints of the real system under investigation. Then, users will select a canonical propagation model that satisfies those conditions and develop a customized network analysis. 
In this Section we show how to construct such families of network analyses. For that, we consider a set of five canonical propagation models---summarized in Fig.~\ref{fig:Figure2}A---representing both discrete and continuous dynamical systems, being also either conservative or non-conservative. 
For each model, we derive their corresponding response matrices: $\mathcal{R}_t$ in the time-discrete cases and $\mathcal{R}(t)$ for time-continuous models. In this scenario, network analysis consists of extracting information out of the $\mathcal{R}$ for the chosen canonical in the form of spatio-temporal network metrics.

\subsection{Discrete and conservative propagation}

A popular propagation model often employed to explore networks is the random walk. The random walk is, as the cascade, a discrete dynamical model both in variable and time. The main difference is that the random walks corresponds to a conservative system in which walkers (or agents), perpetually navigate through the network. In other words, the initial number of walkers is preserved at all times. Given an adjacency matrix $A$, the transition probability matrix $T$ is defined by normalizing the columns by their out-degree, $T_{ij} = A_{ij} / \sum_{k=1}^n A_{kj}$. The entries $T_{ij} \in [0,1]$ thus represent the probability that a walker located at node $j$ at time $t$ will visit one of the (output) neighbors of $j$ at time $t+1$. Formally, the random walker is a mapping $f: \mathbb{R}^n \to \mathbb{R}^n$ of the form:
\begin{equation} 	\label{eq:RandomWalk}
	\mathbf{x}_t = T \, \mathbf{x}_{t-1},
\end{equation}
where $\mathbf{x}_t$ is a column-vector with $x_{i,t}$ representing the expected number of walkers on node $i$ at time $t$. As for the discrete cascade, Eq.~(\ref{eq:RandomWalk}) is solved iteratively. Given initial conditions $\mathbf{x}_0$, then $\mathbf{x}_t = T^t \, \mathbf{x}_0$ for discrete times $t \geq 0$. Initializing the process with one walker at each node, $\mathbf{x}_0 = \mathbf{1}$, the resulting response matrices are $\mathcal{R}_t = \{ T^0, T^1, T^2, T^3, \ldots, T^t \}$.

The first two rows of Fig.~\ref{fig:Figure2}B show the response matrices for the sample graph of $n=8$ nodes depicted, in the case of the discrete cascade and random walks. As seen, the two models give rise to different patterns of pair-wise responses. At the first iteration ($t=1$) the $\mathcal{R}_1$ matrices of both models display a similar pattern that reflects the direct links. But in the subsequent iterations the response matrices begin to differ between the two models. 

The panels on the right display the solutions $x_{i,t}$ for the eight nodes over time. As seen, the number of particles on each node rapidly grows for the cascade while the expected number of walkers on a node stabilizes after a short transient. Additionally, we define the global network response as the sum of all pair-wise responses at each time step, $r(t) = \sum_{i,j=1}^n \mathcal{R}_{ij}(t)$. Figure~\ref{fig:Figure2}C shows the evolution of the corresponding global network responses. For the discrete cascade $r(t)$ rapidly grows but for the random walk model $r(t) = 8$ at all times, corresponding to the eight walkers initially seeded, one per node.

\subsection{Continuous and non-conservative models} 

The extension of Eq.~(\ref{eq:Cascade1}) into continuous time is given by the following differential equation:
\begin{equation} 	\label{eq:ContCascade}
	\dot{\mathbf{x}}(t) = A \, \mathbf{x}(t),	
\end{equation}
where $\mathbf{x}^T(t) = [ x_1(t), x_2(t), \ldots, x_n(t) ]$ is the real-valued (column) state vector of the $n$ variables and $A$ is a real-valued, positive connectivity matrix not restricted to binary. Given initial conditions $\mathbf{x}_0$, the solution of this system is:
\begin{equation}
	\mathbf{x}(t) = e^{At} \mathbf{x}_0. 
\end{equation}
In this case, the pair-wise responses are a continuous function of time
\begin{equation}  	\label{eq:RespCascade}
	\mathcal{R}(t) = e^{At} \, .
\end{equation}
In the engineering and the physics jargons, the response function $\mathcal{R}(t)$ is also known as the Green's function or the propagator. At each time $t$, $\mathcal{R}(t)$ is a matrix of shape $n \times n$ whose elements $\mathcal{R}_{ij}(t) = \left( e^{At} \right)_{ij}$ represent the temporal evolution of the response of node $i$ at times $t > 0$, due to a unit perturbation applied in $j$ at time $t=0$. Equivalently, $\mathcal{R}_{ij}(t)$ can be interpreted as the influence that node $j$ exerts on $i$ after a time lag $t$. As in the discrete cascade, $\mathcal{R}(t)$ encompasses this influence along all possible paths, of all lengths converging into $i$ at different times. Hence, the response is typically larger between nodes sharing a direct connection and smaller between nodes connected via indirect paths. 

The third row of Fig.~\ref{fig:Figure2}B shows the evolution of the response matrices $\mathcal{R}(t) = e^{At}$  for the small sample graph at various temporal snapshots. As seen in the first snapshot, short after the perturbation the responses are governed by the direct connections and $\mathcal{R}(t)$ resembles the adjacency matrix $A$. But as time passes and the influence between nodes expands to longer paths, the pattern of $\mathcal{R}(t)$ changes and it dissociates from $A$. The continuous cascade is also a divergent system and the solutions (node activity) $x_i(t)$ grow exponentially as depicted in the panel at the right. The same trend is observed for the global response $r(t)$, in Fig.~\ref{fig:Figure2}C. \\

Divergent dynamics are rarely representative of empirical systems. A manner to avoid divergence in Eq.~(\ref{eq:ContCascade}) is to include a local dissipative term such that:
\begin{equation} 	\label{eq:ContLeaky}
	\mathbf{x}(t) = - \frac{1}{\tau} \mathbf{x}(t) \, + \, A \, \mathbf{x}(t).
\end{equation}
The term $-\mathbf{x} / \tau$ causes that a fraction of the activity flowing through a node will leak, compensating for the exponential growth of the cascading term $A \mathbf{x}$. The relaxation time-constant $\tau$ controls for the ratio of the leakage: the shorter the $\tau$ the faster the nodes leak. When $\tau \to 0$ the flow is lost through and no activity propagates through the network. Given that $\lambda_{max}$ is the spectral radius of the (weighted) connectivity $A$, the leakage can only compensate the cascading term as long as $0 \leq \tau \leq 1 \,/\, \lambda_{max}$. When $\tau > 1 \,/\, \lambda_{max}$ the exponential growth dominates and the system becomes divergent.

For this model, we define the network response to an initial unit perturbation $\mathbf{x}_0 = \mathbf{1}$ as~\cite{Gilson_DynCom_2018, Gilson_DynComfMRI_2019}:
\begin{equation}  	\label{eq:RespLeaky}
	\mathcal{R}(t) = \left( e^{Jt} - e^{J^0t} \right) \, ,
\end{equation}
where $J = -\delta_{ij} \,/\, \tau + A_{ij}$ is the Jacobian matrix of System~(\ref{eq:ContLeaky}) and $J^0 = - \mathbf{x} \,/\, \tau$ represents the passive leakage of the initial perturbation on a node through itself. In other words, $e^{J^0t}$ are the temporal responses we would observe in an empty graph of $n$ nodes with no links. We choose to regress this out in order to emphasize the part of the responses that are specifically associated to the interactions between nodes. The patterns of the response matrices at the initial times are dominated by the shape of the connectivity matrix $A$, fourth row of Fig.~\ref{fig:Figure2}B. The temporal evolution of the solutions $x_i(t)$ monotonically decay and relax to zero as expected, right panel in Fig.~\ref{fig:Figure2}B. However, the overall network response $r(t)$ undergoes a transient peak in the beginning followed by a decay as the activity of the nodes vanishes, Fig.~\ref{fig:Figure2}C. This behavior results from the interplay of the two terms in Eq.~(\ref{eq:RespLeaky}), each acting at different time-scales. The initial growth is governed by the cascade term and the later, slower decay is controlled by the leakage term.

\subsection{Continuous and conservative models: diffusive coupling and the heat equation}

The two continuous models described so far are non-conservative because the coupling $j \to i$ is mediated by passing the state $x_j$ of (source) node $j$ to (target) node $i$. Thus, the evolution of $i$ depends on the state of its neighbors, which are summed $\sum_{k=1}^n A_{ik} x_k$. In some systems, however, the strength of the interaction between nodes is mediated by their difference $(x_j - x_i)$. For example, this is the case for the Kuramoto model in which the interaction between two oscillators depends on their phase differences: $\dot{\theta}_i \propto \sum_{k=1}^n \sin(\theta_k - \theta_i)$. This type of coupling is termed \emph{diffusive coupling} and is characteristic of conservative dynamical systems. The effect of this interaction is to pull the nodes towards each other, therefore, helping the collective behavior to converge towards a mean-field state.

The simplest linear propagation model based on diffusive coupling is:
\begin{equation} 	\label{eq:Laplacian1}
	\dot{x}_i = \sum_{k=1}^n A_{ik} \left( x_k - x_i\right) \,.
\end{equation}
This expression can be separated as $\dot{x}_i = \sum_{k=1}^n A_{ik} \, x_k - x_i \sum_{k=1}^n A_{ik}$. Since the degree of a node is $d_i = \sum_{k=1}^n A_{ik}$, the expression can be re-written to:
\begin{equation} 	\label{eq:Laplacian2}
	\dot{x}_i = - d_i \, x_i +  \sum_{k=1}^n A_{ik} x_k  \, .
\end{equation}
Defining $D$ as the diagonal matrix with entries $D_{ii} = d_i$, the matrix form of the system is:
\begin{equation} 	\label{eq:Laplacian3}
	\dot{\mathbf{x}} = - D \mathbf{x} + A \mathbf{x} = L \mathbf{x} \, ,
\end{equation}
where $L = -D+A$ is the Jacobian matrix of the linear conservative System~(\ref{eq:Laplacian1}). This Jacobian $L$ is usually named as the \emph{graph Laplacian} in the literature. Comparing Eqs.~(\ref{eq:ContLeaky}) and (\ref{eq:Laplacian2}), it evidences that this conservative system is in fact a special case of the leaky-cascade where the time-constants of the nodes are individually tuned such that $\tau_i = 1 / d_i$. In other words, the input and the leakage ratios of every node are balanced preserving the flow of activity over time.

An alternative but intuitive manner to understand Eqs.~(\ref{eq:Laplacian1})~--~(\ref{eq:Laplacian3}) is to remind that they result as the reduction of the heat equation to graphs (see Appendix~3 and Fig.~\ref{fig:Figure5}). Consider the sample graph of $n=8$ nodes in Fig.~\ref{fig:Figure2}B and imagine that $x_i(t)$ represent the temperature of the nodes. If all nodes were at the same temperature, say $x_i(0) = 5^\circ$C, then no differences exist between adjacent nodes, $A_{ik} \, (x_k - x_i) = 0 \textrm{ for all } \{i,k\}$, and no heat flows. Now, if the temperature of the first node is suddenly raised to $x_1 = 15^\circ$C, heat will start to flow towards its adjacent neighbors $i=2$ and $i=3$, decreasing $x_1$ and increasing $x_2$ and $x_3$. This rebalancing process will extend to the rest of the network through the neighbors of $i=2$ and $i=3$. The temperature $x_1$ will decrease and all other temperatures will increase until all nodes reach $x_i = 6.25^\circ$C, which is in fact the average temperature of the initial conditions $\mathbf{x}_0^T = \left(15, 5, 5, \ldots, 5 \right)$. 

The solution of Eq.~(\ref{eq:Laplacian3}) with initial conditions $\mathbf{x}_0$ is given by $\mathbf{x}(t) = e^{Lt} \, \mathbf{x}_0$. The matrix $e^{Lt}$ is the Green's function (also called the \emph{heat kernel} in some works) of Eq.~(\ref{eq:Laplacian3}). The elements $\left( e^{Lt} \right)_{ij}$ represent the temporal evolution of the response (or influence) of node $i$ at times $t >0$, to a perturbation of unit amplitude in $j$ at $t=0$. In the heat analogy, $\left( e^{Lt} \right)_{ij}$ are the temporal evolution of the temperature at node $i$, assuming that the temperature $x_j$ was suddenly increased by one degree at time $t=0$. As for the case of the leaky-cascade, we regress out the passive leakage term $L^0 = -D \mathbf{x}$ to define the response function:
\begin{equation}  	\label{eq:RespLaplacian}
	\mathcal{R}(t) = \left( e^{Lt} - e^{L^0t} \right) \, .
\end{equation}
For some applications the reader may be interested in directly using $\mathcal{R}(t) = e^{Lt}$, as it has been the case in past examples, see Sec.~\ref{sec:5}, or dividing~\cite{Vilegas_NetRenorm_2022} the contribution of $L^0$ over time. The temporal evolution of $\mathcal{R}(t)$ matrix is depicted in the last row of Fig.~\ref{fig:Figure2}B. As we have observed for other models, in the beginning $\mathcal{R}(t)$ seems governed by the direct connections although at subsequent times the pattern changes and becomes dissociated from the shape of $A$. In particular, the connections of the hub ($i=4$) rapidly loose relevance, as compared to the responses of the leaky-cascade in which the hub is reinforced at early times. Since the initial conditions $\mathbf{x}_0 = \mathbf{1}$ are identical for all nodes, there is no change in the flow across nodes and $x_i(t)$ remain constant over time, right panel at the bottom of Fig.~\ref{fig:Figure2}B. For the same reason, the network responses $r(t) = 8$, which is the sum of the initial unit inputs at the eight nodes.\\

\subsection{Observing networks in the light of different propagation models}

The response matrices depicted in Fig.~\ref{fig:Figure2}B for the five models display rather different patterns of interactions over time despite that, in all cases, they were estimated on the same sample graph. For example, the leaky-cascade seems to emphasize the connections of the central hub ($i=4$) consistently over time, and those of node $i=6$ to a lesser extent. However, in the views of the other models the interactions of the hub $i=4$ are only visible initially but become indistinguishable with time; specially for the random walks and for the continuous cascade. In the case of the continuous diffusive model, the strongest interactions in the intermediate times are some rather peripheral connections: between the first node and its two neighbors, $1-2$ and $1-3$, and the peripheral connections of node $j=6$, $6-7$ and $6-8$. Also, it is notable that the discrete and the continuous cascades provide very distinct patterns at the later times despite both models are divergent.

These observations demonstrate that networks express different views when observed through different propagation models. Beyond the initial time-steps, the patterns of $\mathcal{R}(t)$ always dissociate from the shape of $A$, as the interaction between nodes becomes mediated by longer paths. The resulting patterns at the later times differ across models. Each model tends to highlight some aspects of the connectivity and shadows other aspects. These observations strengthen the idea that network structure cannot always be abstracted away from network dynamics and shall be jointly studied. It also underlines that, in order to carry out interpretable network analyses, it is necessary to choose a propagation model that respects the minimal constraints of the real system under investigation.

\section{Examples and Applications}		\label{sec:4}

Section~\ref{sec:3} presented five canonical models to perform model-based network analyses. We now provide examples to illustrate that this dynamical approach to study networks can overcome some limitations of classical graph analysis. We first deal with the problem of network comparison and then we explain how to derive a more general definition of distance between nodes based on the time-scales of propagation. The section ends with the presentation of analytical results to provide formal support to these observations. For illustrative reasons we restrict the calculations to the leaky-cascade model and its response functions, Eq.~(\ref{eq:RespLeaky}), but the derivation are generalizable to the other models.

\subsection{Comparing networks with each other}

\begin{figure*}[hp!]  
	\centering
	\includegraphics[width=1.0\textwidth,clip=]{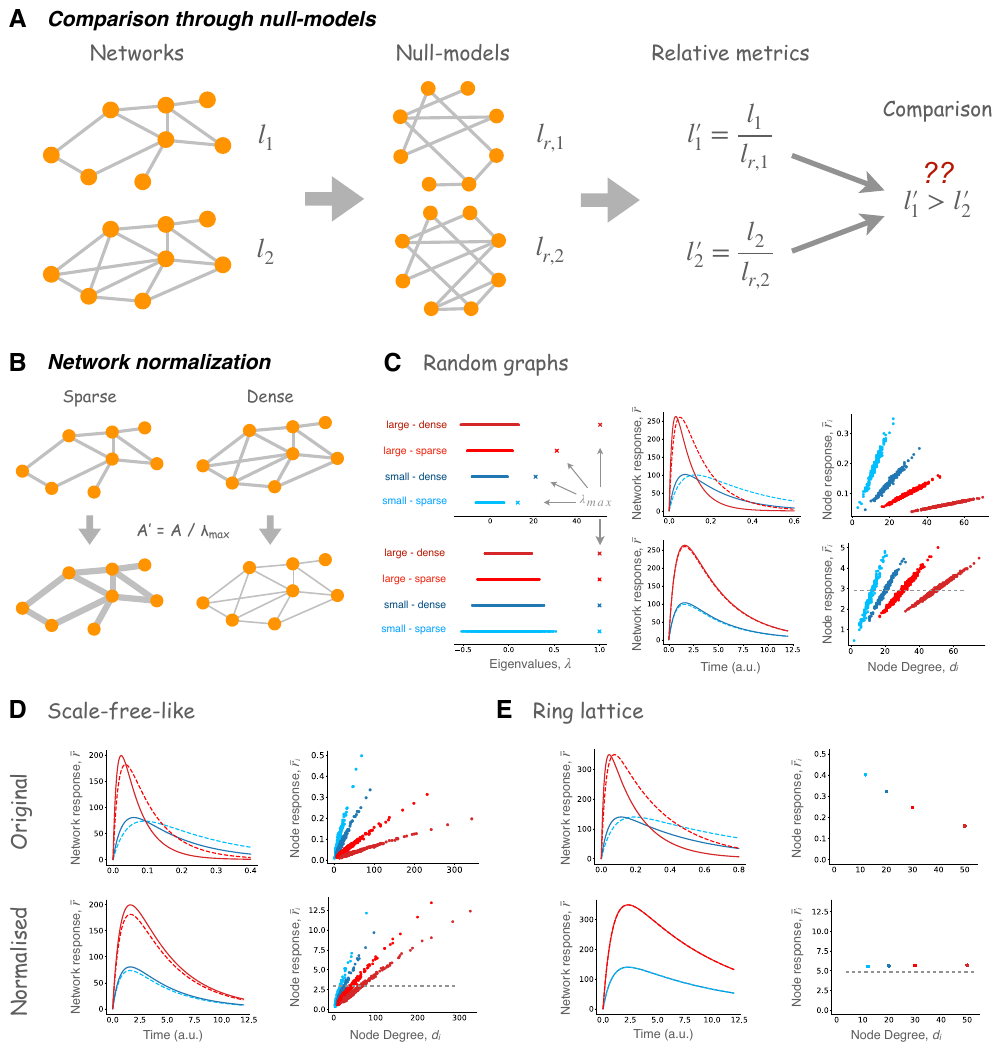}
 	\caption{		\label{fig:Figure3}
	{\bf Normalizing connectivity to compare networks.}
	{\bf A} Schematic representation of classical approach in graph analysis to compare networks via third-party comparison to null-models. {\bf B} Schematic representation of network comparison in the dynamical perspective. Normalizing the connectivity matrices by their largest eigenvalues $\lambda_{max}$ gives rise to weighted connectivities with aligned time-scales of evolution.
	{\bf C} -- {\bf E} Testing network comparison by renormalization of networks with largest eigenvalue for three familiar network models (random graphs of uniform probability, scale-free-like random graphs and ring lattices). Top panels display results out of the original (binary) adjacency matrices and bottom panels show results for the normalized connectivities. For each model, four distinct graphs were generated, from small and sparse to large and dense, combining sizes $n_1 = 200$ and $n_2 = 500$ nodes with densities $\rho_1 = 0.06$ and $\rho_2 = 0.1$. Scale-free-like graphs were generated using degree-degree probabilities that would return an exponent $\gamma = 3.0$ in the thermodynamical limit. The time-constant $\tau$ for each network was set such that $\tau = 0.8 \, \tau_{max}$, where $\tau_{max} = 1 \,/\,\lambda_{max}$. 
	} 
\end{figure*}

The outcome of graph metrics is influenced by the size $n$ and the density $\rho$ (or number of links $m$) of a network. This dependence makes it difficult to compare across networks. Imagine we study two graphs $G_1$ and $G_2$ of same size $n_1 = n_2$ but one is denser than the other, say $\rho_1 = 0.01 < \rho_2 = 0.06$. If we obtained average pathlengths $l_1 = 3.5$ and $l_2 = 2.9$ respectively, this means that $G_2$ is shorter than $G_1$. However, it is well-known that the pathlength decays with the number of links. Additionally, we may also want to ask whether $l_2 < l_1$ only because $G_2$ is denser than $G_1$, or because their internal architecture differs. In order to answer this question we would need to regress out the influences of both size and density on the pathlength, which is not always possible. Therefore, the typical strategy to deal with this problem is to compare empirical networks to simple graph models (null-models)~\cite{Molloy_Config,Newman_Config_2001,Humphries_SWness_2008,Barthelemy_RandomGeometric_2018,Vasa_NullModels_2022}, e.g. random graphs or degree-conserving random graphs. As depicted in Fig.~\ref{fig:Figure3}A, we would construct two ensembles of random graphs of matching $n$ and $m$ of $G_1$ and $G_2$, and we would compute the ensemble average pathlengths $l_{r,1}$ and $l_{r,2}$. Then, we would compare the relative metrics $l'_1 = l_1 \,/\, l_{r,1}$ and $l'_2 = l_2 \,/\, l_{r,2}$ with each other to derive conclusions about which network architecture is shorter.

This typical procedure suffers from some conceptual limitations because random networks only offer relative comparisons, instead of absolute extremal values~\cite{Zamora_Sizing_2019}. Nevertheless, from the dynamical perspective to network analysis here proposed there is no need to employ null-models for comparing networks. Instead, a simple normalization of the connectivity suffices to align networks of different size and/or density, making them comparable. The largest eigenvalue of a connectivity matrix $\lambda_{max}$ captures the intrinsic time scale of a network for linear dynamical models. Hence, for any two networks, normalizing the connectivity matrices by their corresponding $\lambda_{max}$ such that $A' = A \,/\,\lambda_{max}$, aligns their responses making the two networks directly comparable~\cite{Zamora_Hubs_2010,Arnaudon_Centrality_2020}; see Fig.~\ref{fig:Figure3}B. The largest eigenvalues of the normalized connectivities $A'_1$ and $A'_2$ are the same: $\lambda'_{max,1} = \lambda'_{max,2} = 1.0$. It shall be noted that after the normalization the matrices $A'_1$ and $A'_2$ are weighted. Standard graph theory cannot deal with these normalized connectivities as it requires adjacency matrices to be binary, with entries 0 or 1. However, this dynamical approach to network analysis naturally deals with such weighted networks. 

In Figs.~\ref{fig:Figure3}C-E we show the results of this normalization on three network models: random graphs (uniform probability), scale-free-like graphs and ring-lattices. For each of the three models we generate four graphs combining sizes $n = 200$ or $500$, and densities $\rho = 0.06$ or $0.1$. We study the responses $\mathcal{R}(t)$ of the networks using the continuous leaky-cascade. For each type of network, the global network responses $r(t)$ of the four graphs (top panels) follow different amplitudes and characteristic time-scales despite their internal architecture is equivalent---as they are instances of the same graph model. Next, we normalized the connectivity matrices by their corresponding $\lambda_{max}$ and recomputed the responses $\mathcal{R}'(t)$. As shown (bottom panels), the normalization aligns the response curves $r'(t)$ collapsing them in pairs of different amplitude. The difference in response amplitudes depend only on the network size. 
A further normalization of the responses by network size $n$ would fully align the four curves. However, this would not necessarily make the networks more comparable since the $\lambda_{max}$ normalization implies that the total response per node is the same in all networks. 

We illustrate this studying the relation between the node-wise responses and the node degrees in the original graph before and after the normalization. The node responses $r_i(t)$ are defined as the temporal response of a node to all the initial perturbations. It is thus computed as the row sum of the response matrices: $r_i(t) = \sum_{k=1}^n \mathcal{R}_{ik}(t)$. Then, the total node response $\bar{r}_i$ accounts for the accumulated response at the node from the initial time $t=0$. It is calculated as the integral (or area under the curve) over time $\bar{r}_i = \int_{t=0}^\infty r_i(t) dt$. A linear relation between the original degrees $d_i$ of the binary graph and the total node responses $\bar{r}_i$ is observed in all the networks. Networks generated from the same graph model follow the same degree distribution but the actual values $d_i$ grow with $n$ and $\rho$. In the comparison to $\bar{r}_i$ before the normalization (top panels), we find the same trend happens for the $\bar{r}_i$ values taken by the nodes. Their absolute values grow with $n$ and $\rho$ of the underlying original graphs forming separate ``clouds'' of points in the plots. However, after the adjacency matrices have been normalized (bottom panels), the values for the responses $\bar{r}'_i$ of the four networks become aligned, showing that both the $\bar{r}_i$ values and their distributions $p(\bar{r}_i)$ are now directly comparable across networks.

\subsection{Defining graph distance from response times}

\begin{figure*}[ht!]
	\centering
	\includegraphics[width=1.0\textwidth,clip=]{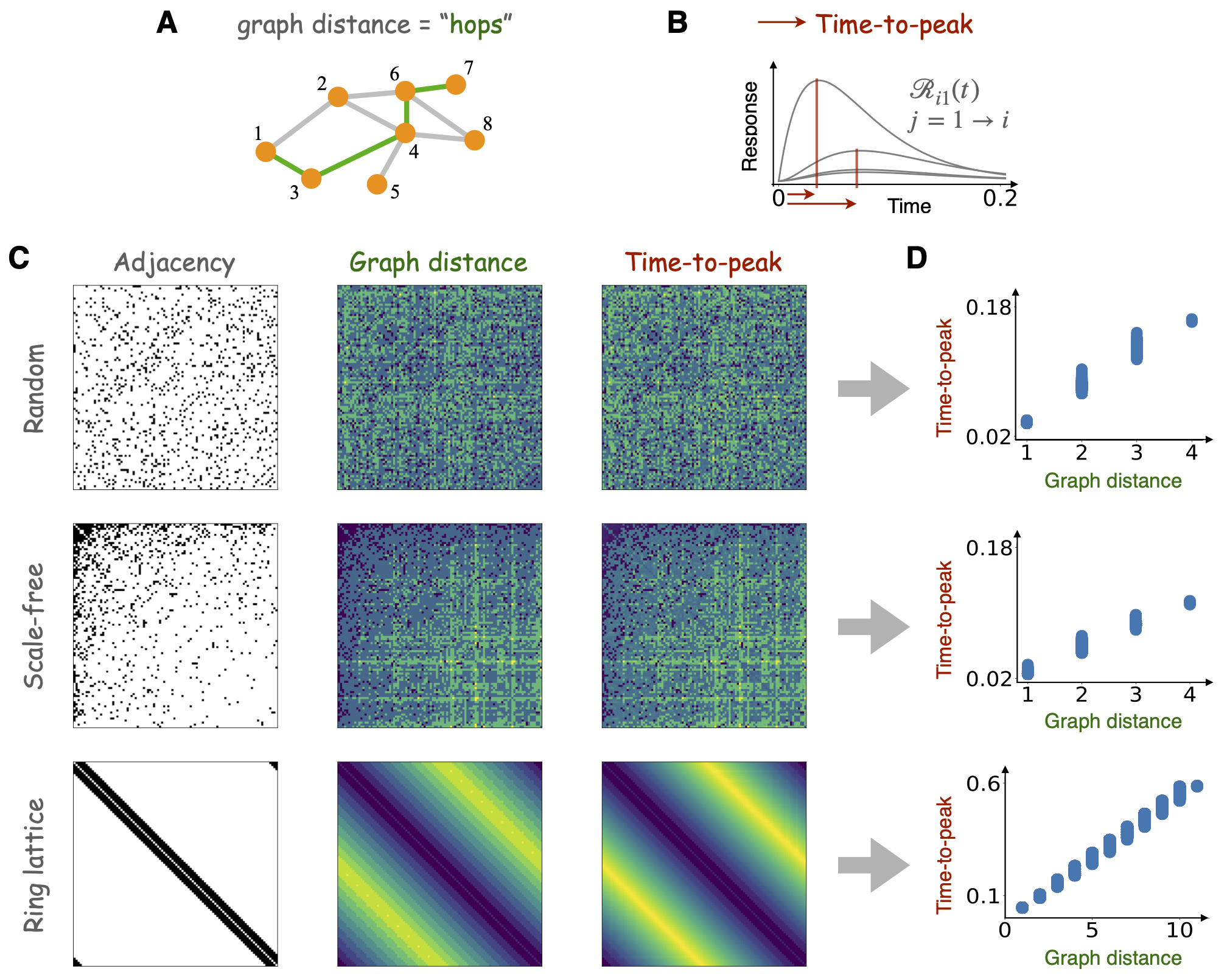}
 	\caption{		\label{fig:Figure4}
	{\bf Redefinition of distance in networks based on response times.}
	{\bf A} In graphs, the distance between two nodes is defined as the number of links needed to traverse to travel from one node to another. {\bf B} Under the dynamical perspective here proposed, the distance can be redefined in terms of the time needed for a unit stimulus at $j$ (column index, source node) to take effect on $i$ (row index, target node). This could be quantified in different manners, e.g., the time it takes for the response to reach its peak value. {\bf C} and {\bf D}, validation of equivalence between classical graph distance and time-to-peak distance in three sample graphs: a random graph, an scale-free graph and a ring lattice. {\bf C} Adjacency matrices for the three sample graphs and the corresponding pair-wise distance matrices following the two approaches. Qualitatively, both distance matrices look very much the same. {\bf D} Comparison of the distance between all pairs of nodes in the three graphs computed either as classical graph distance $D^g_{ij}$ or time-to-peak $D^{ttp}_{ij}$. The relation is linear---corroborating the agreement between the two measures---although some degeneracy in the time-to-peak is found across pairs that lie at same graph distance.
	} 
\end{figure*}

In graphs, the (geodesic) distance $D^g_{ij}$ between two nodes is defined as the smallest number of links that an agent needs to traverse---hopping from node to node---to travel from source $j$ to target $i$, Fig.~\ref{fig:Figure4}A. However, this concept is only valid for the case of unweighted, binary graphs as its calculation relies on discrete agents or particles navigating through the network. If the  edges of a graph are weighted, or the system is represented by continuous variables, the geodesic distance cannot be computed. In Appendix~1 we show that from the dynamical perspective, the graph distance between two nodes corresponds to the discrete time step $t$ at which a discrete cascade initiated at node $j$ arrives for the first time at node $i$. This redefinition of distance in terms of time allows for a more flexible generalization, applicable to other propagation models. 

In the leaky-cascade, the pair-wise responses $\mathcal{R}_{ij}(t)$---describing the response of $i$ to an initial perturbation at $j$---undergo a transient growth followed by a slower decay, as depicted in Fig.~\ref{fig:Figure4}B. In this scenario, we can define the distance from $j$ to $i$ as the time required for the response $\mathcal{R}_{ij}(t)$ to reach the peak value. We illustrate this time-to-peak distance $D^{ttp}_{ij}$ for three undirected graphs (random, scale-free-like and a ring lattice) of $n = 100$ and density $\rho = 0.1$. The scale-free-like network was generated for $\gamma = 2.5$. The adjacency matrices, the graph distance $D^g_{ij}$ and the time-to-peak distance $D^{ttp}_{ij}$ matrices for the three graphs are shown in Fig.~\ref{fig:Figure4}C. Visually, $D^g_{ij}$ and $D^{ttp}_{ij}$ look very much alike, indicating that in these unweighted cases measuring time-to-peak or the classical graph distance are qualitatively equivalent. Quantitatively, the agreement is not exact but similar, Figure~\ref{fig:Figure4}D. While graph distance is a discrete quantity, time-to-peak is continuous. Thus there is some level of degeneracy in the time-to-peak values taken by nodes at the same geodesic distance. Although this variation is small and a reasonable linear correlation is found between the two metrics.

It shall be reminded that in the particular case of the leaky cascade, the response dynamics depends on the intrinsic relaxation time-constant $\tau$ governing the rate of the leakage. For the examples in Fig.~\ref{fig:Figure4}, $\tau$ were independently chosen in the three networks such that $\tau = 0.4 \, \tau_{max}$ with $\tau_{max} = \frac{1}{\lambda_{max}}$. The value of $\tau$ can alter the linear relation between $D^g_{ij}$ and $D^{ttp}_{ij}$ with wider degeneracy and ultimately saturating when $\tau \to \tau_{max}$.

\subsection{Mean-field and link-wise calculations for the leaky-cascade}

We finalize this section providing analytical calculations to support the previous observations. We firstly detail mean-field approximations for the network response $r(t)$, which can provide intuition about the normalization proposed above to compare networks. This normalization is equivalent to rescaling the networks in order to obtain the same mean weight per node in random networks (for which the mean-field approximation is accurate). Note that such mean-field approximations are also useful when the entire calculation of the response matrices $\mathcal{R}(t)$ turn computationally expensive. We then formally proof the above observations on the time-to-peak distance $D^{ttp}_{ij}$. See further details in Appendix~2. Derivations for other models than the leaky-cascade should follow similar rationale, e.g., see Ref.~\cite{Arnaudon_Centrality_2020} for the continuous diffusive model.

The pair-wise responses $\mathcal{R}_{ij}(t)$ from source $j$ to target $i$ can be evaluated using the spectral properties of the adjacency matrix $A$ as follows~\cite{Gilson_DynCom_2018}:
\begin{equation*}
	\mathcal{R}_{ij}(t)=\sum_{k} \left[ e^{-t/\tau+\lambda_{k}t}-e^{-t/\tau} \right] \left( u_{i}^{T}v_{k} \right) \left( u_{j}^{T}v_{k} \right)  \, , 
\end{equation*}
where $\lambda_k$ are the eigenvalues of $A$, $v_k$ their corresponding right eigenvectors and $u_i = \delta_{ij}$ are the unitary vectors. The mean-field approximation consists in evaluating the contribution associated to the largest eigenvalue and discarding all other contributions, while also assuming homogeneity in the network. This implies that the dominating eigenvector is close to the uniform vector and the dominating eigenvalue is $\lambda_{1} \simeq A_{in} = \sum_{i,j} A_{ij} / N(N-1)$. This yields
\begin{equation}
	r(t)  \simeq  e^{-t/\tau} \left( e^{\lambda_{1}t}-1 \right)  \simeq  e^{-t/\tau} \left( e^{A_{in}t}-1 \right) \, ,
\end{equation}
that is, the node-wise metrics obtained when summing over all inputs $r_i(t) = \sum_{k=1}^n \mathcal{R}_{ik}(t)$, as also done with the integral over time $t$ above yielding $\bar{r}_i$. In this way, we get a quantification for node importance (either at time point $t$ or in total), providing a ranking of nodes in the network, as centrality measures do. The key point here is that the distribution of values $\bar{r}_i$ for nodes $i$ can be compared across networks because the (mean-field) baseline has been matched.

We now provide analytical results to confirm the close-to-linear relationship between time-to-peak and graph distances in Fig.~\ref{fig:Figure4}. As shown in Appendix~1, the geodesic distance $D^g_{ij}$ is the smallest exponent of the adjacency matrix $A$ such that the corresponding matrix element is non-zero: 
\begin{equation*}
	D^g_{ij} = \mathrm{argmin}_{l} \left[ (A^{l})_{ij}>0 \right] .
\end{equation*}
On the other hand, by definition, the time-to-peak $t^*$ corresponds to the time-point at which the derivative of the response curve $\mathcal{R}_{ij}(t)$ for the pair $(i,j)$ is zero, leading to the equality $\left[ e^{At} \right]_{ij} = \left[ (I-\tau A)^{-1} \right]_{ij}$, with $I$ being the identity matrix. Using the power series expansions for both sides, the matrix exponential and the inverse matrix, and applying the Stirling approximation of the factorial, we obtain the approximate time-to-peaks:
\begin{equation}
	t^* \simeq \frac{\tau}{e} D^g_{ij}  \, ,
\end{equation}
which implies that the time-to-peak $t^* = D^{ttp}_{ij}$ is globally linearly related to the geodesic distance $D^g_{ij}$, as illustrated in Fig.~\ref{fig:Figure4}D. Moreover, the connection-wise peak amplitude satisfies the following equality:
\begin{equation*}
	\mathcal{R}_{ij}(t^*) = e^{-t^*/\tau} \left( \sum_{n\geq0} \tau^{n} A^{n} \right)_{ij} \propto K_{ij}  \, ,
\end{equation*}
with $\tau$ being the rescaling factor for the adjacency matrix $A$ and the matrix $K = \frac{1}{I - \tau A} $  , with the fraction indicating the matrix inverse, $\left( I - \tau A \right)^{-1}$. This means that the node-wise peak amplitudes $r_i(t^*) = \sum_{k=1}^n \mathcal{R}_{ik}(t^*)$ is proportional to the Katz centrality; see Sec.~\ref{sec:5} and Appendix~2 for further details.

These results confirm that the responses $\mathcal{R}_{ij}(t)$ provide a decomposition of the information necessary to calculate graph metrics that capture network effects in accordance to usual concepts such as distances between nodes and node ranking to describe hierarchies. Again, we stress that aligning the networks by their dominating eigenvalues and then comparing their peak amplitudes would also allow to contrast Katz centrality across networks of different sizes or densities.

\section{Relation with literature of modern and complex network analysis} 		\label{sec:5}

So far, we have shown that classical graph metrics arise from a common propagation model, we have proposed five canonical models to generalize network analyses under different constraints and we have illustrated some benefits of redefining network analysis under a dynamical perspective. This last section reviews recent efforts in the literature to define network measures that are---in one way or another---based on dynamical phenomena. Our aim is to clarify the propagation model, the assumptions and the constraints behind those different metrics. In doing so, we identify that the majority of these methods follow one of the five canonical models in Fig.~\ref{fig:Figure2}. In general, we find two classes of approaches.
On the one hand, we encounter heuristic measures inspired by dynamical rationale but whose underlying propagation dynamics are hidden, implicitly assumed or unknown. Examples of this class are the Katz centrality and communicability. On the other hand, we find methods in which the underlying spreading, navigation or diffusion process is explicit. This class comprises approaches for both discrete units (e.g., random walks) and continuous variables (e.g., diffusion or flows).

Concerns regarding implicit assumptions, hidden dynamics and the lack of transparency are not new in the study of networks, specially regarding the definition of centrality measures: e.g., degree, closeness, betweenness, eigenvector or Katz centralities. Many of these were defined following intuitive but implicit---or even hidden---dynamical motivations. This lack of transparency has motivated debates calling for further clarity in the field~\cite{Friedkin_Centrality_1991, Bogartti_Centrality_2005}. As stated by S.P. Bogartti~\cite{Bogartti_Centrality_2005}:
\begin{quote}
What is not often recognized is that the formulas for these different measures make implicit assumptions about the manner in which things flow in a network (\ldots) the discussion of centrality has largely avoided any mention of the dynamic processes that unfold along the links of a network (not to mention the processes that shape the network structure). Yet, the importance of a node in a network cannot be determined without reference to how traffic flows through the network.
\end{quote}
He concludes that:
\begin{quote}
\ldots the off-the-shelf formulae for centrality measures are fully applicable only for the specific flow processes they are designed for, and that when they are applied to other flow processes they get the `wrong' answer. It is noted that the most commonly used centrality measures are not appropriate for most of the flows we are routinely interested in.
\end{quote}
These concerns resonate with the aim of this Perspective Article although, here, our intention is to generalize these ideas to the essence of network analysis beyond centrality.

\subsection{Heuristic measures of influence: Katz centrality and communicability}

We start by exposing the hidden dynamical origin and the implicit assumptions of two popular network measures: Katz centrality and communicability. We establish the connection between the two measures thanks to the dynamical viewpoint endorsed in this article.

Given the fact that the powers of the adjacency matrix $A^l$ determine the number of paths of length $l$ between nodes, it has been often recognized that this construct should be key to explain the functional relation between them. Accordingly, it has been proposed that the \emph{influence} that one node exerts over another should depend on the accumulated effect through all possible routes, of all lengths, available to travel between the nodes. This idea of influence thus surpasses the contribution of direct connections and shortest paths, and it can be quantified as the following sum:
\begin{equation*}
Q = A + A^2 + A^3 + A^4 + A^5 + \ldots
\end{equation*}
The problem with this expression is that for any binary adjacency matrix the sum diverges as the values $(A^l)_{ij}$ rapidly grow with length $l$. In order to avoid this, Katz (1952)~\cite{Katz_Centrality_1952} proposed to include an attenuation factor $\alpha < 1$ which ``\emph{has the force of a probability of effectiveness of a single link.}'' In other words, $\alpha$ is a weight given to the links which tunes---reduces---the efficiency of transmission. Then, the expression above is rewritten as: 
\begin{equation}	\label{eq:Qsum}
Q = \alpha A + \alpha^2 A^2 + \alpha^3 A^3 + \alpha^4 A^4 + \alpha^5 A^5 + \ldots 	
\end{equation}
When $\alpha$ is small enough, the reduced efficiency of transmission compensates the growth of the powers $(A^l)_{ij}$ and guarantees the convergence of the sum. In fact, convergence is achieved only if $0 < \alpha < 1 \,/\,\lambda_{max}$, where $\lambda_{max}$ is the largest eigenvalue of $A$. In this case, the sum is a power series leading to the exact expression:
\begin{equation}	\label{Eq:QKatz}
Q^K = \mathrm{I} + \alpha A + \alpha^2 A^2 + \ldots = \sum_{l=0}^\infty \left( \alpha A \right)^l   \equiv   \frac{1}{\mathrm{I} - \alpha A} \; ,
\end{equation}
where $\mathrm{I}$ is the identity matrix. Following this, Katz defined the centrality of the nodes $\mathbf{c}^K$ as:
\begin{equation}	\label{eq:CKatz}
\mathbf{c}^K   =  \left( Q^K -I \right) \mathbf{u}   =   \left( \frac{1}{\mathrm{I} - \alpha A} - \mathrm{I} \right) \mathbf{u} \, .
\end{equation}
Here, $\mathbf{u}$ is a (column) vector of input strengths and the matrix $Q^K$ encodes the net \emph{influence} that one node exerts over another through all possible paths of all lengths. If we assume a stimulus of unit amplitude at all nodes, $\mathbf{u}^T = \mathbf{1}^T = (1,1, \ldots, 1)$, then Katz centrality $c_i^K$ quantifies the cumulative influence that these perturbations will have on $i$, excluding the self-influence triggered by the perturbation $u_i = 1$. Katz centrality is sometimes expressed as
\begin{equation}
	\mathbf{c}^K = Q^K \mathbf{u} = \left( \frac{1}{\mathrm{I} - \alpha A} \right) \mathbf{u}
\end{equation}
which includes the self-influences.

A more recent measure to estimate the influence of nodes beyond shortest paths is the communicability metric~\cite{Estrada_Communicability_2008}. In this case, the convergence of the power series is guaranteed by choosing the factorial coefficients $1 / l!$ such that 
$\mathrm{I} + A + \frac{1}{2!} A^2 + \frac{1}{3!} A^3 + \ldots = \sum_{l=0}^{\infty} \frac{A^l}{l!}$.
Algebraically, this series represents the matrix exponential $e^A$ and converges for all positive definite matrices $A$. In some applications an optional factor $\alpha$ is included. Then, communicability is defined as the influence matrix:
\begin{equation} 	\label{eq:QCommun}
	Q^C =  \mathrm{I} + \alpha A + \frac{1}{2!} \alpha^2 A^2 +  \ldots  =  \sum_{l=0}^{\infty} \frac{(\alpha A)^l}{l!}   \equiv   e^{\alpha A}.
\end{equation}

In Katz centrality and communicability the factor $\alpha$ tunes the ``depth'' of the path structure at which influence is exerted. When $\alpha = 0$ no information or influence can pass across the links. Increasing $\alpha$ will first favor transmission along direct connections. Further increase of $\alpha$ gradually allows for longer paths to take effect and---in the case of the Katz centrality---when $\alpha > 1 \,/\,\lambda_{max}$ it will cause $Q^K$ to diverge. Conceptually, the factor $\alpha$ can be interpreted in different manners; either as an attenuation factor, a resistance or more generally, as a coupling strength associated to the links. \\

As seen, the rationale behind Katz centrality and communicability is identical. Both approaches define the (time-averaged) influence that one node exerts over another accounting for the cumulative influence propagated through all possible paths, of all lengths, which are encoded in the power matrices $A^l$. From this algebraic point of view, the only difference between the two approaches is that communicability is biased, favoring the influence of shorter paths while Katz centrality is more sensitive to the longer path structure. In Katz centrality the propagation through a link suffers the same attenuation $\alpha$ at every step, regardless of how many steps were given before. In contrast, the factorial coefficient $1 / l!$ of communicability punishes the longer paths, which in fact facilitates the convergence of the power-series of Eq.~(\ref{eq:QCommun}).

Now, from the dynamical perspective we propose here, we can clarify that the difference between Katz centrality and communicability is that they assume different propagation models. Katz centrality is based on the leaky-cascade of Eq.~(\ref{eq:ContLeaky}) and communicability assumes the continuous cascade in Eq.~(\ref{eq:ContCascade}).
 
Considering the leaky cascade subjected to a constant external unit input $\mathbf{u} = \mathbf{1}$, Eq.~(\ref{eq:ContLeaky}) is written as $\dot{\mathbf{x}} = - \mathbf{x} / \tau \, + \, A \mathbf{x} \, + \, \mathbf{1}$. The steady-state solution ($\dot{\tilde{\mathbf{x}}} = \mathbf{0}$) is given by $\left( \mathrm{I} / \tau - A \right) \tilde{\mathbf{x}} = \mathbf{1}$. Solving for $\tilde{\mathbf{x}}$ we have that~\cite{Tononi_Complexity_1994,Galan_DominatingPatterns_2008,Zamora_FComplexity_2016,Sharkey_Katz_2017}:
\begin{equation} 	
	\tilde{\mathbf{x}} = \left( \frac{1}{\mathrm{I} / \tau - A} \right) \mathbf{1},
\end{equation}
which is proportional to the definition of Katz centrality ($\tilde{\mathbf{x}} = \tau \, \mathbf{c}^K$) in the version that accounts for the self-influences and with attenuation factor $\alpha = \tau$. Whether $\alpha$ divides the leakage term such that  $\dot{\mathbf{x}} = -\mathbf{x} / \alpha + A \mathbf{x}$ or multiplies the connectivity matrix such that $\dot{\mathbf{x}} = -\mathbf{x} \,+\, \alpha A$ is just a matter of convenience. the parameter $\alpha$ can be interpreted accordingly either as a leakage time-constant, a coupling strength or a dissipation factor but, mathematically, those forms are identical.
Regarding communicability, we recall that the solution of the parametrized continuous cascade $\dot{\mathbf{x}}(t) = \alpha A \, \mathbf{x}(t)$ to initial conditions $\mathbf{x}_0$ is given by $\mathbf{x}(t) = e^{\alpha At} \, \mathbf{x}_0$. From here, it is trivial to realize that communicability $Q^C = e^{\alpha A}$ is a special case of the pair-wise response (or Green's) function $\mathcal{R}(t) = e^{\alpha At}$ at a single temporal snapshot, when $t = 1$. That is, communicability $Q^C \equiv \mathcal{R}(t=1)$.

A benefit of the dynamical perspective against the algebraic one is that the relation between the responses (quantifying influence) and the length of the paths is naturally unfolded in the temporal dimension. An input at $t=0$ is first transmitted through the shorter paths and influence of longer paths takes effect at later times.

\subsection{Describing networks through random walks}

Due to their Markovian (no memory) and conservative nature, random walks on networks are mathematically tractable and well-behaved (the dynamics do not diverge). Hence, they are a convenient tool to explore networks and to describe them according to the emerging patterns while walkers flow through a network. Specially relevant have been their applications for community detection~\cite{Pons_Walktrap_2006,Rosvall_Infomap_2008,Delvenne_StabilityComms_2010,Piccardi_LumpedMarkov_2011,Schaub_Encoding_2012} and for defining centrality measures~\cite{Page_PageRank_1998, Newman_Betweenness_2005}.

The notion of a community (module or cluster) in a network is that of a subset of nodes more densely interconnected with each other than they are with the rest of the network. The core idea behind community detection with random walks is that walkers are trapped into communities: they spend more time wandering within the communities than jumping between them. In practice none of the algorithms follows the temporal evolution of a walker. Instead, they employ estimates of the within-module and cross-modular transition probabilities, either at the long-time horizon---the steady-state solution---or at a given time step.

For example, the \emph{Netwalk} algorithm~\cite{Zhou_Brownian_2003, Zhou_Dissimilarity_2003, Zhou_Netwalk_2004} evaluates the time-horizons at which nodes \emph{see} other other. For each node $j$, it first estimates the mean-first-passage-time $D^1_{ij}$ to all other nodes. The mean first-passage time is defined as the average number of steps that a walker starting at $j$ takes to visit $i$ for the first time. Then, the algorithm clusters together nodes with similar distance profiles (columns of the $D^1_{ij}$ matrix). The idea is that two nodes in the same community should reach other nodes within a similar time horizon.

The \emph{Walktrap} algorithm~\cite{Pons_Walktrap_2006} directly employs the powers of the transition probability matrices $T^t$. Although, instead of using all the information in the response matrices $\mathcal{R}_t = \left\{ T^0, T^1, T^2, T^3, \ldots \right\}$ as we suggested in Sec.~\ref{sec:3}, Walktrap focuses at one intermediate time $t'$ arbitrarily chosen ``\ldots \emph{long enough to gather enough information about the topology of the graph (but) must not be too long (compared to the mixing time of the Markov chain), to avoid reaching the stationary distribution.}'' Once a time $t'$ is selected, the algorithm clusters together nodes that have similar (outgoing) transition probability profiles---the columns of $T^{t'}$. The idea is that two walkers that start from the same community should have similar probability of being found at a given node at time $t'$. \\

The centrality of a node can be regarded as its tendency to attract walkers. Since random walks are Markovian, in the long-time horizon a walker has no memory of where it started. Therefore, random walk centrality $c^{rw}_i$ is usually quantified as the steady-state solutions $\tilde{x}_i$. For the simple random walk in Eq.~(\ref{eq:RandomWalk}), the probability of a walker to visit a node at the steady-state is trivially proportional to the degree of the node (its input degree if the graph is directed). Hence, in this case $c^{rw}_i = \frac{d_i}{\sum_{k=1}^n d_k} = \, ^1 \!/ _N \, \tilde{x}_i$, where $N$ is the number of walkers seeded at $t=0$. Defining centrality as the steady-state solution misses the temporal transient information but allows to employ the same criteria for different classes of random walks. 

PageRank~\cite{Page_PageRank_1998} is another popular measure of random walk centrality. At its core we find the special case of random walks with teleportation. In this model a walker navigates through network as in the simple random walk but from time-to-time, with probability $(1-\epsilon)$, it jumps---teleports---to any of the $n$ nodes of the network. In large networks, e.g., the world-wide web, the teleportation term allows the walkers to explore a local vicinity of the network before jumping to a different region. The time spent in a vicinity is thus controlled by $\epsilon$. 
This process is governed by the iterative equation:
\begin{equation}  	\label{eq:WalkerTeleport}
	\mathbf{x}_t  = \left[(1-\epsilon) \mathbf{v} +  \epsilon \, T  \right] \, \mathbf{x}_{t-1} \, ,
\end{equation}
where $\mathbf{v}$ is a preference vector encoding the probability $v_i$ that the walker jumps to node $i$ in case of teleportation. If this is uniform, then $v_i = \, ^1\! / n$. But otherwise it allows to specify a ranking of preferences if we had such information about the real system. PageRank centrality $c^{PR}_i$ is calculated as the steady-state solution $\tilde{x}_i$ of Eq.~(\ref{eq:WalkerTeleport}). Mathematically, this corresponds to the right eigenvector of the transition matrix $T_{PR} =  \left[ (1-\epsilon) \mathbf{v} + \epsilon \, T \right]$ that is associated with the largest eigenvalue $\lambda_{max} = 1$ (the \emph{left} eigenvector if we followed the graph convention $i \to j$).

\subsection{Network metrics based on continuous diffusion}

Several network metrics have been proposed in the literature based on continuous diffusion. For example, through the definition of 
diffusion kernels~\cite{Kondor_Kernels_2002}. This approach follows a similar reasoning as for Katz centrality and communicability. Diffusion kernels define an influence matrix between nodes which is then employed to derive the metrics. Zhang and coauthors proposed a heuristic Gaussian kernel~\cite{Zhang_Mapping_2010}, $K_{ij} = \exp \left( D_{ij} / 2h^2 \right)$, which defines the influence between nodes depending on their shortest-path distance $D_{ij}$. The parameter $h$ modulates the depth of the interactions along distances, playing a similar role as the attenuation factor $\alpha$ in Katz centrality and communicability. This kernel has been employed to define a centrality measure~\cite{Zhang_NodeImportance_2011} and to study communities~\cite{Zhang_KermelsCommunity_2009} at different levels of resolution, with $h$ controlling for the resolution scale.

Another example of heuristic kernels for diffusion is the redefinition of communicability that replaces the adjacency matrix $A$ by the graph Laplacian $L$ such that~\cite{Estrada_Review_2012}:
\begin{equation}
	Q^L =  \mathrm{I} + \alpha L + \frac{1}{2!} \alpha^2 L^2 + \frac{1}{3!} \alpha^3 L^3 +  \ldots   \equiv e^{\alpha L}  \, .
\end{equation}
While the original communicability arises from the continuous cascade $\dot{\mathbf{x}} = \alpha A \, \mathbf{x}$ as we showed before, it is easy to identify now that $Q^L$ arises from the continuous diffusion $\dot{\mathbf{x}} = \alpha L \, \mathbf{x}$. In fact, $Q^L$ is the Green's function of this equation at time $t=1$ with parameter $\alpha$ modulating the influence of the longer paths. Therefore, the different between the two is that $Q^C$ follows a divergent linear propagation and $Q^L$ a conservative systems with the nodes diffusively coupled as $(x_k - x_i)$.

A recent study~\cite{Arnaudon_Centrality_2020} proposed a measure of centrality that scans across different hierarchical scales making use of the Green's function $e^{Lt}$ of the continuous diffusion model applied to the graph Laplacian. Recalling that $\left( e^{L t} \right)_{ij}$ is the temporal response of node $i$ to the unit stimuli at $j$, the authors defined the distance between two nodes as the time $t^*_{ij}$ at which the curve $\left( e^{L t} \right)_{ij}$ peaks. This allows to explore the centrality of nodes at various temporal horizons, and reveal different scales of the network. We note that the rationale behind this centrality measure is the same as the definition of time-to-peak distance discussed in Sec.~\ref{sec:4}, only that for the examples in Fig.~\ref{fig:Figure4} we employed the leaky-cascade of Eq.~(\ref{eq:ContLeaky}).

Finally, several methods have been proposed to identify communities across different resolution scales based on continuous diffusion on networks~\cite{Cheng_CommunityDiffusion_2010,Delvenne_StabilityComms_2010,Schaub_Encoding_2012,Lambiotte_RandomWalks_2014}. In principle, the reasoning behind these methods is the same as those for random walks but employing a time-continuous formulation. In these cases, the goodness of a partition is evaluated in terms of the temporal stability of the communities---say, the average time that a walker seeded in one community at $t=0$ would need to leave the community. Paying attention to the transient times $t$ (the Markov time) before the steady-state horizon ($0 < t < t_{\infty}$), it allows to zoom-in and zoom-out the resolution and identify communities of different size. The time-continuous system associated to the simple random walk is given by:
\begin{equation} 	\label{eq:RWcont}
	\dot{\mathbf{p}}  = - \mathbf{p} + T \, \mathbf{p} =  [- I + T] \, \mathbf{p} \, .
\end{equation}
Recalling that the transition matrix is calculated as $T = D^{-1} A$, Eq.~(\ref{eq:RWcont}) reduces to:
\begin{equation} 	\label{eq:RWcont2}
	\dot{\mathbf{p}} =  [-I + D^{-1} A] \, \mathbf{p} = \left( D^{-1} L \right) \, \mathbf{p} = \mathcal{L} \, \mathbf{p} \, ,
\end{equation}
where $\mathcal{L} \equiv D^{-1}L = -I + T$ is the (column) normalized Laplacian matrix. Expressed in terms of the individual elements we have:
\begin{equation} 	\label{eq:RWcont3}
	\dot{p}_i = - p_i + \sum_{k=1}^n T_{ik} p_k = \ldots = \frac{1}{d_i} \sum_{k=1}^n A_{ik} \left(p_k - p_i \right)  \, .
\end{equation}
From here we see that, despite these methods were motivated by random walks, Eqs.~(\ref{eq:RWcont2}) and (\ref{eq:RWcont3}) are the same as the continuous diffusion Eqs.~(\ref{eq:Laplacian2}) and (\ref{eq:Laplacian3}) only that the adjacency matrix $A$ is replaced by the transition matrix $T$. The values of $\mathcal{L}$ are bounded to $[0, 1]$ so that $p_i(t)$ represent the temporal evolution of probabilities. This is the reason for why we summarized these methods here instead of together with methods for random walks. Besides, it shall be noted that despite the analogies drawn in the literature, the solutions $\mathbf{x}_t$ for the time-discrete random walk in Eq.~(\ref{eq:RandomWalk}) and the time-continuous version $\mathbf{p}(t)$ in Eq.~(\ref{eq:RWcont}) converge only at the steady-state but do not follow the same time-courses along the transition, see Appendix~4 and Fig.~\ref{fig:Figure6}. 
Therefore, the time-discrete and the time-continuous random walk lead to different views of a network which turns the same only in the $t\to \infty$ limit.  \\

At this point, we find it useful to clarify that both communicability $Q^L $, the hierarchical centrality measure of Ref.~\cite{Arnaudon_Centrality_2020} and the Markov-time community detection methods~\cite{Cheng_CommunityDiffusion_2010,Delvenne_StabilityComms_2010}, are all constructed upon the same model: the canonical continuous diffusion of Eqs.~(\ref{eq:Laplacian1})-(\ref{eq:Laplacian3}). Despite they were introduced following different reasoning and motivations: heuristically, from its analogy to the heat equation or as a time-continuous version of random walks. This level of degeneracy in the literature makes it difficult for users to navigate through the abundance of methods. It is in part for this reason that in the present Article we advocate for more transparency such that all network measures explicitly and clearly state the canonical model underlying each of them. This clarity can only help users make better methodological choices and derive more accurate interpretations of the results obtained.

\subsection{An integrative perspective for model-based network analyses with propagation dynamics}

In this Section, we have summarized network metrics that were proposed in the past, which were inspired by the idea of probing propagation dynamics on networks to characterize them. We have exposed their underlying propagation models to highlight their differences and similarities. It shall be noted that all these methods are valid measures to describe networks. However, given that they are bound to different assumptions and constraints, it is important to understand when it makes sense to use one method or another, and how their results should be interpreted. Making the right choice will depend on the information we have about the real system, or the hypotheses we may want to test. Rather than competing with each other, all these measures form a family of complementary methods for studying networks. 

The idea that different dynamical processes could be exploited to extract complementary views out of a network has been explored in the past. For example, Zhang et al.\cite{Zhang_Mapping_2010} compared the outcome of propagation kernels, phase oscillators, epidemic models and random walks on a set of empirical and synthetic graphs. Regarding community detection Lambiotte et al. (2014)\cite{Lambiotte_RandomWalks_2014} suggested that the simple random walk could be replaced by other types of random walks, e.g. biased random walks, with teleportation or with memory:
\begin{quote}``\ldots the choice of (random walk) dynamics can also be used to find the most appropriate community structure (if particular information about the system is available) or to explore the network under different (and complementary) viewpoints to gain deeper information about the system.''
\end{quote}
In this Article, our aim has been to generalize these ideas ($i$) providing a perspective that embraces the diversity of real systems studied as networks, and ($ii$) to propose a unified methodology to deal with the ecosystem of network metrics. Specifically, we propose that network analyses should follow these steps:
\begin{enumerate}[leftmargin=12pt,]
\item Identify the {\bf key constraints} of the real system. For example, the user may need to question whether the system investigated is discrete or continuous, whether it is conservative or non-conservative, whether it follows divergent or convergent behavior, whether it is diffusively coupled or not.
\item Choose a corresponding {\bf canonical propagation model}. We proposed the five canonical models shown in Fig.~\ref{fig:Figure2} which, in turn, are the root of many metrics in the literature as we just reviewed. But users may consider other canonical models if required by the constraints of the real system. For example, we saw that PageRank assumes a special case of random walks with teleportation.
\item Compute the pair-wise {\bf network responses} $\mathcal{R}(t)$ of the network, for the chosen canonical model. Using the Green's function $\mathcal{R}(t)$ of the system is a natural choice for linear propagation models, but further study is necessary for an extension to non-linear models. Note that, in our definitions for the leaky-cascade and the continuous diffusion models, we regressed out the passive leakages $J^0$ and $L^0$ in order to highlight the contributions of the responses due to the interactions.
\item Extract information out of $\mathcal{R}(t)$ in the form of {\bf spatio-temporal network metrics}. For example, here we defined the global network response $r(t)$, the time-to-peak distance $D^{ttp}$ for the leaky-cascade and the node-wise responses $r_i(t)$. Several other metrics are possible depending on the questions one may have about the real system. In general, both $\mathcal{R}_{ij}(t)$, $r(t)$ and $r_i(t)$ are the curves of the temporal evolution of the responses, and a variety of information could be extracted from those. Also relevant are the self-responses $\mathcal{R}_{ii}(t)$ which generalize the graph concepts of node degree and clustering coefficient, but at different times. Community detection methods could take advantage of the input / output response profiles of the nodes---the rows and columns of $\mathcal{R}(t)$---in order to find communities at different times of resolution. 
\end{enumerate}

\section{Discussion and future perspectives}

The goal of this Perspective Article is to motivate a paradigm shift in the analysis of complex networks, from a data-driven (model-free) tradition to a model-based culture. Graph theory has been considered a data-driven analysis tool and therefore, its metrics applicable to any system that is represented as a graph. As reviewed in Sec.~\ref{sec:5}, many other measures have been proposed in the literature to characterize networks beyond graph theory, in particular for centrality and community detection. We have exposed that all these methods---including classical graph metrics---are either heuristically or explicitly based on canonical propagation models. We derive two lessons from here. First, network analysis is already model-based only that a consequent tradition is still missing. Corresponding analysis practices and a unified formulation that encompasses the different model-based metrics were still needed. And second, the fact that all past methods were defined upon different dynamical models, subjected to particular assumptions and constraints, it begs to question the presupposed universality of network metrics. Despite they were introduced as if they were useful to study any network.

Here, we have proposed a unified formulation to harmonize the network analyses that use different propagation models. We envision that in the future, network analyses should begin by choosing an adequate canonical model---not necessarily restricted to the five suggested in Sec.~\ref{sec:3}. Then, one should estimate the (pair-wise) network response function $\mathcal{R}(t)$ for the corresponding model in the network of interest. And finally, we would extract the information about the network from $\mathcal{R}(t)$ in the form of spatio-temporal network metrics. Although defining such metrics might not always be trivial, we have shown that this dynamical point of view to network analysis brings several benefits. 

We acknowledge that the transition from data-driven to model-based analysis can only happen at the cost universality---a price many will find difficult to pay given the traditions in the field. But by doing so, there is plenty of specificity and interpretability to gain. 
The success of PageRank as a website ranking tool is a prominent example. The reason for why PageRank outperformed other algorithms is because its underlying propagation model---the random walk with teleportation in Eq.~(\ref{eq:WalkerTeleport})---is a very simplistic but reasonable account of human behavior when navigating through the world-wide web. Usually, a person navigates the web clicking on hyperlinks found on the current webpage  until the person ``looses interest and jumps'' to restart the navigation visiting websites of a different topic. The success of PageRank thus lies in the fact that its underlying propagation model shares minimal but necessary ingredients of the real system it aims at measuring. For the same reason, it is unlikely that PageRank could be a good centrality measure for other real networks, e.g., neural networks, epidemic spreading, traffic or protein-protein interactions, for which the assumptions and constraints of a conservative random walk with teleportation do not make sense.

One issue with model-based data analyses is that of model selection, which may lead to controversies between authors with different views on which is \emph{the adequate} model for each case. Another potential source of dispute may lie in the distinction between canonical models (intended for the analysis) and actual models (meant to reproduce the system). The difference between analysis and modeling can sometimes become a gray area with no clear boundaries. However, we are convinced that the transparency and the interpretability of results that a model-based approach brings are beneficial for the field of complex network. And, in our opinion, it could even be seen as a sign of its maturity. For example, in statistics, one would never apply certain metrics to a dataset unless the data passes a Gaussianity test before. Because if the data is not normally distributed, the numerical results computed out of those measures would not be interpretable. We believe that it is better for the field of complex networks to deal with these controversies than to continue without a  ``Gaussianity test,'' trapped in the dream of universality. 

Before the study of a network, users should perform a careful selection and identify which metrics make sense for the real system under investigation. And which metrics should be discarded. But to do so, users need clear and transparent information about the assumptions behind each metric.
For example, neurons that communicate via chemical (pulse) coupling transmit spikes (discrete units) which propagate and multiply through the network, sometimes giving rise to avalanches. Therefore, it is unreasonable to study neuronal and brain networks with metrics based on random walks or in the graph Laplacian, simply because the brain is not a conservative system and neurons are not diffusively coupled.

We shall also emphasize that the dynamical perspective supported here is neither to be regarded as \emph{the only} solution to network analysis. Surely there are real-world systems susceptible of a graph representation but whose study in terms of propagation and navigation is not meaningful. Their characterization may require employing other forms of analyses and different metrics. The key lesson is that, as users, we should choose the right tools for each case. And as method developers, we should transparently inform of the assumptions, the constraints and the range of validity of the tools we provide.

\subsection{Limitations and future work}

Our proposal to derive network metrics out of the pair-wise response functions $\mathcal{R}(t)$ as spatio-temporal measures comes with a notable limitation. Numerical calculation of the $\mathcal{R}(t)$ tensors for a large number of time steps could be computationally expensive, and prohibitive for large networks. In those cases, analytical (mean-field) approximations as the ones derived in Sec.~\ref{sec:3} and Appendix~2 could become very useful. Also, as accounted in past approaches, one may need to restrict the analysis to the steady-state solutions. Although these options miss the richness of detail allowed by our proposal, they will still benefit from the desired transparency of model selection.

Despite we proposed to derive network metrics out of $\mathcal{R}(t)$, here we did not provide a complete set of possible measures. The redefinition of distance in networks~\cite{Arnaudon_Centrality_2020} and the global network response metrics are useful starting points. In previous applications to neuroimaging data (employing the leaky-cascade) we calculated the input and the output responses to each node, which are reminiscent of the integrative and broadcasting capacities of the brain regions~\cite{Gilson_DynComfMRI_2019,Panda_IntegBroadcast_2023}. Surely, users will also derive further measures out of $\mathcal{R}(t)$, depending on the information they may like to extract from their networks. Additionally, we find it rather interesting the possibility to generalize the classical graph metrics, which are bounded to discrete times $t = 1-3$, to longer time-scales or under other canonical models.

In the present Perspective we did not directly treat the case of weighted or directed networks. We restricted the examples to binary and undirected graphs in order to undoubtedly state the mapping from graph theory to its dynamical representation. But the application to directed and weighted networks is straight forward. For example, in the comparisons of Sec.~\ref{sec:4} the normalized networks are all weighted. Also, we studied empirical weighted networks in past publications following this formulation~\cite{Gilson_DynCom_2018,Gilson_DynComfMRI_2019,Panda_IntegBroadcast_2023}.
Despite this dynamical approach can naturally deal with weighted networks, we shall emphasize again that the interpretability of the results is bounded to those cases where the weights of the real system are associated to physical or statistical notions that are compatible with the models of propagation and diffusion.

Finally, the work presented in this Perspective Article is tightly related to the broader problem of the structure-function relation in networks. We omitted such discussion to focus on the implications for data analysis alone. The relation between network structure and dynamics has been vastly studied in the literature. It is helpful to frame those efforts into three categories: ($i$) works trying to investigate the effect of a known architecture on the collective dynamics. ($ii$) Efforts aiming at describing the architecture of a network by probing dynamics on them. This is the main focus of the present Article. And ($iii$) attempts to infer the unknown links of a network by use of empirically observed dynamics. 

It is well-known in the literature that the collective dynamics on a network not only depend on the connectivity but also on the dynamical model we assume, leading to different views of the same network~\cite{Gomez_EntropyRate_2008,Zhang_Mapping_2010,Hens_SignalPropag_2019,Ji_PropagationReview_2023} as we illustrated in Fig.~\ref{fig:Figure2}. Some of these works also aim at describing the propagation of perturbations on networks for different models~\cite{Barzel_MinimalModels_2015,Harush_PatternsFlows_2017,Bao_BasicMotifs_2022}. However, in our opinion, the literature should deal with this information with more precision, more stratified. In some occasions, the goals of papers do not seem well classified within the three categories just mentioned. In other occasions, it may even seem that a single paper intends to solve all the problems associated to the three cases at once. We can only speculate about the reasons. On the one hand, due to the lack of a tradition for model-based network analyses we may be prone to overlook the difference between canonical models and actual models, instead of exploiting their distinctive purposes separately. On the other hand, the desire---or the community pressure---to deliver universal results could make it difficult for authors to frame their work more specifically. We can only hope that our efforts here to seed some clarity, and our call for more transparency in the field, may help the community to contextualize future contributions to the network-dynamics problem with more precision.

\acknowledgements{
This work has been supported (GZL and MG) by the European Union's Horizon 2020 research and innovation programme under Specific Grant Agreement No. 785907 (HBP SGA2) and Specific Grant Agreement No. 945539 (Human Brain Project SGA3).
MG acknowledges funding from French Agence Nationale de la Recherche (ANR) and AMidex (Aix-Marseille University).
} 

\section*{Appendix 1: Dynamical representation of graph metrics}	\label{sec:Appendix1}

From the point of view of graph theory, all the relevant information about the network is encoded in the adjacency matrix $A$. Combinatorial or algorithmic methods allow then to answer different questions about the architecture of the graph in the form of graph metrics. The dynamical paradigm shown in the previous sections exposes that under the discrete cascade in Eq.~(\ref{eq:Cascade1}), the structural information in $A$ is unfolded into the set of powers $\mathcal{R} = \{ A^0, A^1, A^2, A^3, \ldots, A^t \}$ representing the temporal response of the network to initial stimuli applied in at nodes. We now show how fundamental graph metrics are encoded by the response matrices $\mathcal{R}$.

The \emph{degree} of a node, $d$, is defined as the number of neighbors of the node. Assuming undirected graphs with $A$ symmetric, it is calculated as the row or column sum of the adjacency matrix such that $d_i = \sum_{k=1}^n A_{ik}$. In the dynamical perspective, the degree is expressed as the number of particles returning to the node in the short time scales. A particle starting at node $j$ produces that each neighbor of $i$ receives one particle in the first iteration. In the second iteration, $t=2$, new particles will propagate to the neighbors of each node. The single particle starting from $j$ at $t=0$ results in $j$ receiving one particle per neighbor at time $t=2$. In other words, the degree is the influence that a node exerts on itself at time $t=2$ and it is thus represented by the diagonal elements of $A^2$:
\begin{equation}
	d_i = (A^2)_{ii}. 		\label{eq:Degree}
\end{equation}

\emph{Matching index} is a measure of the structural similarity between two nodes. It is evaluated counting the number of common neighbors since, two nodes that share the same connections play an identical role in the graph. Given that $\mathcal{N}(v)$ is the set of nodes connected to vertex $v$---the neighborhood of $v$---the number of common neighbors between two nodes is quantified as the size of the overlap of their neighborhoods: $m(i,j) = |\mathcal{N}(i) \cap \mathcal{N}(j)|$. From the adjacency matrix $m(i,j)$ is calculated comparing the columns corresponding to the two nodes such that $m(i,j) = \sum_{k=1}^n A_{ik} A_{jk}$. Usually, the matching index $M(i,j)$ is normalized by the fraction between the number of common neighbors and the total number of nodes adjacent to either $i$ or $j$:
\begin{equation} 	\label{eq:Matching1}
	M(i,j) = \frac{|\mathcal{N}(i) \cap \mathcal{N}(j)|}{|\mathcal{N}(i) \cup \mathcal{N}(j)|}
		 = \frac{m(i,j)}{d_i + d_j - m(i,j)}.
\end{equation}
Thus, $M(i,j) = 0$ when $i$ and $j$ have no neighbors in common and $M(i,j) = 1$ when both nodes are connected to the same, and only the same, neighbors.

Under the perspective of the discrete cascading, the overlap $m(i,j)$ can be regarded as the ``convergence zone'' of two simultaneous propagations, one starting from $i$ and the other from $j$. Imagine the initial conditions $\mathbf{x}_0$ with $x_{k,0} = 1$ if $k = i,j$ and $x_{k,0} = 0$ otherwise. After the first iteration, nodes adjacent to either $i$ or $j$ will receive one particle and the only nodes with two particles are those adjacent to both $i$ and $j$.
At the second time-step, node $j$ receives one particle, due to the initial one on $i$ at $t=0$, from each of the nodes shared with $i$. Therefore, the number of common neighbors $m(i,j)$ between $i$ and $j$ is reflected in the matrix element $(A^2)_{ij}$. In other words, the influence that $i$ exerts on $j$ at time $t=2$---or the influence of $j$ on $i$---is mediated exclusively via their common neighbors. If they had no common neighbors, then there is no influence between them at this time step. As shown before, the degrees $d_i$ are encoded in the entries $(A^2)_{ii}$, thus substituting in Eq.~(\ref{eq:Matching1}) we can express the matching index in terms of the discrete propagation as:
\begin{equation} 	\label{eq:Matching2}
	M(i,j) = \frac{(A^2)_{ij}}{(A^2)_{ii} + (A^2)_{jj} - (A^2)_{ij}},	
\end{equation}
This expression illustrates that in dynamical terms the normalized matching index is regarded as the fraction of the influence between two nodes that is routed via the common neighbors, and it thus invites for a generalization of the index to subsequent time steps $t > 2$ by allowing the subsequent powers into Eq.~(\ref{eq:Matching2}). Such a generalization should also open the door to define an equivalent metric when the underlying discrete cascade is replaced by other more general dynamical models.

The \emph{clustering coefficient}, $C$, is a popular graph metric. It quantifies the probability that the neighbors of one node are connected with each other. In social terms, it answers the question of how likely is that ``my friends are also friends with each other''. In practice, the clustering coefficient is calculated by counting the number of triangles in a graph since a link between two neighbors of a node leads to a triangle. It is well-known that the diagonal entries of $A^3$ represent the number of triangles---cycles of length $l=3$---in which nodes participate and that the total number of triangles in a graph is given by $n(\bigtriangleup) = \frac{1}{3} \, tr(A^3) = \frac{1}{3} \sum_{i=1}^n (A^3)_{ii}$, where the factor $\frac{1}{3}$ is to account for the fact that every triangle is counted once per node. For the clustering coefficient to be a probability, it is normalized by the total number of triads $n(\vee)$, or paths of length $l = 2$ in the graph. Thus, $C$ is 1 only if all the triads form closed paths. In terms of the powers of $A$, the total number of paths of length $l = 2$ is calculated as $n(\vee) = |A^2| - tr(A^2)$, where $| \cdot |$ represents the sum of all the elements of the matrix, and $tr(\cdot)$ is the trace. So, the clustering coefficient is calculated as:
\begin{equation}
	C = 3 \, \frac{n(\bigtriangleup)}{n(\vee)}  = \frac{tr(A^3)}{|A^2| - tr(A^2)}.
\end{equation}
Under the dynamical perspective of the discrete cascading in Eq.~(\ref{eq:Cascade1}), the quantity $|A^2| - tr(A^2)$ represents the number of particles that are generated in the iteration from $t=1$ to $t=2$, or in other words, the total influence exerted across nodes at time $t=2$. The quantity $tr(A^3)$ is the number of particles returning to the nodes, or the self-influence at $t=3$. Thus, in dynamical terms the clustering coefficient can be interpreted as how much of the influence generated by the network at time $t=2$ falls back to the nodes at $t=3$. 

It shall be noted that if the degree is a metric of the influence of a node over itself at $t=2$, the clustering coefficient is a metric of the influence that nodes exert on themselves at time $t=3$. The difference is that the clustering is normalized in order to take the form of a probability. Last, the dynamical definition of $C$ allows for a natural generalization of the probability of self-interaction at any time, such that for all $t >1$,
\begin{equation}
	C_t  = \frac{tr(A^t)}{|A^{t-1}| - tr(A^{t-1})}.
\end{equation}

The \emph{geodesic distance}, $D_{ij}$, between two nodes in a graph is defined as the minimal number of links needed to traverse in order to reach $i$ from $j$, or pathlength. Graph distance cannot be derived from the adjacency matrix alone. Its calculation requires to navigate through the graph, e.g. based upon DFS or BFS algorithms. As mentioned before, the cascading process described in Eq.~(\ref{eq:Cascade1}) is indeed the BFS navigation without memory. Under this cascading, instead of counting the number of jumps to travel between nodes, $D_{ij}$ can be evaluated from a temporal point of view---as the time needed for a cascade initialized at node $j$ to reach node $i$ for the first time. That is, in dynamical terms graph distance can be regarded as the time a perturbation on a node needs to reach the other nodes, instead of the ``static'' definition of the pathlength. Given the set of matrix powers $\mathcal{R} = \{ A^0, A^1, A^2, A^3, \ldots, A^t \}$, we can formally redefine graph distance as:
\begin{equation} 	\label{eq:Distance}
	D_{ij} = t' : \: (A^{t'})_{ij} > 0 \; \textsf{if for all} \; t < t', (A^t)_{ij} = 0.
\end{equation}
Our overall goal is to generalize graph analysis by replacing the original propagation dynamical model behind graph metrics, Eq.~(\ref{eq:Cascade1}), with other models which account for other basic properties of real systems. The interpretation of distance in terms of the time required for perturbations to propagate will become a handful change of perspective under arbitrary dynamical rules, either discrete or continuous.

\section*{Appendix 2: Analytical calculations}	 	\label{sec:Appendix2}

\paragraph{Responses for individual pairs of nodes reflect the spectral properties of the adjacency matrix.}

We can express the matrix elements $R_{ij}(t)$ in terms of the eigenvalues $\lambda_{k}$ and the corresponding (right) eigenvectors $v_{k}$ of the adjacency matrix $A$. For that, we use the property of matrix $\mathbf{v}=(v_{k})$ whose columns are the orthogonal eigenvectors satisfying $\mathbf{v} \, \mathbf{v}^T=1$. Using the column unitary vectors $\mathbf{u}$ with entries $u_{i}=\delta_{ik}$. For the case in which the response matrices are defined as $R(t)=e^{At}$, we have:
\begin{eqnarray}
R_{ij}(t)  &  =   &   u_{i}^{T}R(t)u_{j} = u_{i}^{T}e^{At} \mathbf{v}\,\mathbf{v}^T u_{j}	\nonumber  \\
              &  =   &  \sum_{k}(u_{i}^{T}e^{At}v_{k})(v_{k}^{T}u_{j})   				\nonumber \\
              &  =   &  \sum_{k}e^{\lambda_{k}t}(u_{i}^{T}v_{k})(u_{j}^{T}v_{k})	\, ,		\nonumber 
\end{eqnarray}
which is a weighted sum of exponentials of the eigenvalues (multiplied by the time). Note that $u_{i}^{T}v_{k}$ is simply the $i$-th coordinate of $v_{k}$. For the responses of the discrete cascade $R(t)=A^{t}$, we have similarly
\[
R_{ij}(t)=\sum_{k}\lambda_{k}^{t}(u_{i}^{T}v_{k})(u_{j}^{T}v_{k}) \, ,
\]
and for the case $R(t)=e^{Jt}-e^{J^{0}t}=e^{-t/\tau}(e^{At}-\boldsymbol{1})$,
\[
R_{ij}(t)=\sum_{k}e^{-t/\tau}(e^{\lambda_{k}t}-1)(u_{i}^{T}v_{k})(u_{j}^{T}v_{k}) \, .
\]
The mean field approximation consists in discarding all eigenvalues aside from the dominating one, say $\lambda_{1}\simeq A_{in}$ and approximating the corresponding eigenvector with the uniform vector $\bar{u}=\sum_{i}u_{i}$. For the three cases, this returns 
\begin{displaymath}
\bar{R}(t) \simeq 
	\left\{
	\begin{array}{l}
	e^{\lambda_{1}t}\simeq e^{A_{in}t}  	\\
	\lambda_{1}^{t}\simeq A_{in}^{t}		\\
	e^{-t/\tau}(e^{\lambda_{1}t}-1)\simeq e^{-t/\tau}(e^{A_{in}t}-1)  \, .\\
	\end{array}
	\right.
\end{displaymath}

\paragraph{Time-to-peak (TTP) is linearly related to the geodesic distance.}

If we consider a single pair of nodes, from source $j$ to target $i$, for the case of the leaky-cascade, then TTP can be calculated using the derivative of $R(t)\propto e^{Jt}-e^{J^{0}t}$, which is 
\[
\frac{1}{\tau}e^{-t/\tau}(e^{At}-1)=Ae^{-t/\tau}e^{At}
\]
yielding the condition $e^{At}=(1-\tau A)^{-1}$. In principle, we would need to evaluate the logarithm of the matrices but we look for an approximation. Let's call $p$ the geodesic distance for connection from $i$ to $j$ that verifies $p=\mathrm{argmin}_{l} \left[ (A^{l})_{ij}>0 \right]$. That is, the smallest exponent of $A$ such that the corresponding matrix element is non-zero, see Appendix~1. Then, TTP $t$ is defined as the solution of $e^{At}=(1-\tau A)^{-1}$, which can be expressed using the series expansions for the matrix exponent in both sides:
\[
\sum_{n\geq0}\frac{t^{n}A^{n}}{n!}=\sum_{n\geq0}\tau^{n}A^{n} \, .
\]
For all integers $0\leq n\leq p-1$, the terms are zero and the first non zero term is for $n=p$. Because of the factorial in the denominator of the left side, and of the power of $\tau$ on the right side, this first term dominates on each side meaning that in fact
\[
\frac{t^{p}A_{ij}^{p}}{p!}\simeq\tau^{p}A_{ij}^{p}
\]
and $A_{ij}^{p}=1$ disappears. To approximate the factorial, we can use the Stirling approximation of the factorial $p!\simeq\sqrt{2\pi p}\left(\frac{p}{e}\right)^{p}$ that implies $(p!)^{(1/p)}\simeq\frac{p}{e}$ (even though it is designed for large $p$, it is a reasonable approximation here sincewe pass it to the p-order root thereafter), so the previous equality can be rewritten as $\frac{t^* e}{p} \simeq \tau$ which means that TTP is linearly related to the geodesic distance by $t^* \simeq \frac{\tau}{e}p$.

\paragraph{Sum of peak amplitudes (SPA) for a node is linearly related to Katz centrality.}

At the peak, we have the derivative condition satisfied for the matrix element corresponding to the link $[e^{At^*}]{}_{ij}=[(1-\tau A)^{-1}]_{ij}$, so the amplitude is given by
\begin{eqnarray}
R_{ij}(t^*)	&  =  &  \left[e^{Jt^*}-e^{J^{0}t^*}\right]_{ij}  =  e^{-t^*/\tau} \left[e^{At^*}-1\right]_{ij}	\nonumber	\\  	
		&  =  &  e^{-t\*/\tau} \left[e^{At^*}\right]_{ij}  =   e^{-t^*/\tau} \left(1-\tau A \right)_{ij}^{-1}		\nonumber
\end{eqnarray}
using $i\neq j$. It can be further expressed using the same series as before:
\[
R_{ij}(t^*)=e^{-1/\tau} \left[ \sum_{n\geq0}\tau^{n}A^{n} \right]_{ij} \, .
\]
It follows that the peak amplitude takes the form of the Katz centrality, connection-wise. The node-wise measure would just be obtained summing over $j$), with attenuation parameter $alpha$ taking the role of the leakage constant $\tau$ (and the overall expression rescaled by a factor).

\section*{Appendix 3: From heat equation to the graph Laplacian}	\label{sec:Appendix3}

\begin{figure*}[ht!]
	\centering
	\includegraphics[width=1.0\textwidth,clip=]{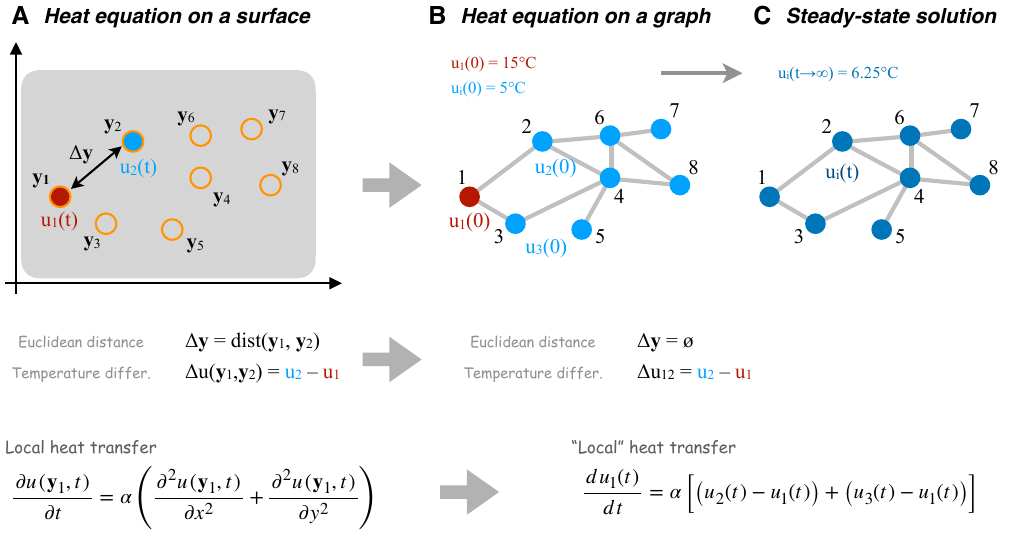}
 	\caption{		\label{fig:Figure5}
	{\bf The heat equation for graphs: origin of the graph Laplacian.}
	The heat equation describes the heat transfer in a continuous media, which depends on the difference in temperature between two points and on their euclidean distance. Given the fact that graphs are discrete objects with no spatial embedding, euclidean distance between nodes is not defined. Instead, the ``local vicinity'' of a node $i$ consists of those nodes $k$ with which it shares a connection. Therefore, the local heat transfer between nodes depends uniquely on their difference in temperature $u_k - u_i$. In consequence, the Laplacian operator $\nabla^2$ of the heat equation for continuous media, $\dot{u}(y,t) =  \alpha \nabla^2 u(\mathbf{y},t)$, turns the Laplacian matrix $L$ on a discrete graph such that $\dot{\mathbf{u}} = \alpha L \mathbf{u}$. 
	} 
\end{figure*}

An intuitive manner to understand Eqs.~(\ref{eq:Laplacian1})~--~(\ref{eq:Laplacian3}) is to remind that they are the translation of the heat equation to graphs. Imagine a two dimensional surface as in Fig.~\ref{fig:Figure5}A. The heat transfer between two nearby points at the surface, $\mathbf{y}_1$ and $\mathbf{y}_2$, is proportional to the gradient of temperature between them. This gradient depends both on the difference in temperature $\Delta u_{12}(t) = u(\mathbf{y}_2,t) - u(\mathbf{y}_1,t)$ and on the euclidean distance between the two points $\Delta \mathbf{y} = \parallel \mathbf{y}_2 - \mathbf{y}_1 \parallel$. The heat transfer $u(\mathbf{y},t)$ at one point of the surface is thus evaluated as the local gradient $ du(\mathbf{y},t) \propto \left( u(\mathbf{y}^*,t) - u(\mathbf{y},t) \right)$ between the point $\mathbf{y}$ and all other points $\mathbf{y}^* = \mathbf{y} + d\mathbf{y}$ that are infinitesimally close to it. In this case, the heat equation at point $\mathbf{y}$ is written as:
\begin{equation} 	\label{eq:Heat2D}
	\frac{\partial u(\mathbf{y},t)}{\partial t} =
	\alpha \nabla^2 u(\mathbf{y},t) =
	\alpha \left( 
	\frac{\partial^2 u(\mathbf{y},t)}{\partial x^2} + \frac{\partial^2 u(\mathbf{y},t)}{\partial y^2}
	\right) \, ,
\end{equation}
where $\alpha > 0$ is a diffusivity parameter and $\nabla^2$ is the Laplacian operator that evaluates the local gradient around a point in an Euclidean space.

The heat equation defines diffusion in a continuous media. But graphs are discrete objects without a spatial embedding, Fig.~\ref{fig:Figure5}B. Therefore, the ``local vicinity'' of a node $i$ consists of the set of nodes with which $i$ shares a link. If we assume that nodes have temperature $u_i(t)$, the gradient of temperature around $i$ simply becomes the sum of the differences $\sum_{k=1}^n A_{ik} \left(u_k - u_i \right)$ because, in a graph, there is no spatial dimension and thus, there is no euclidean distance between them. From here, it is rather simple to realize that Eqs.~(\ref{eq:Laplacian1})~--~(\ref{eq:Laplacian3}) are equivalent to (\ref{eq:Heat2D}) with $\alpha = 1$ and the Laplacian operator $\nabla^2$ for the local gradient taking the form of the Laplacian matrix $L$. Given that the equation governing the temperature of node $i$ is $\dot{u}_i = \sum_{k=1}^n A_{ik} \left(u_k - u_i \right)$, this expression can be separated as $\dot{u}_i = \sum_{k=1}^n A_{ik} u_k \,-\, \sum_{k=1}^n A_{ik} u_i$. Reminding that $d_i = \sum_{k=1}^n A_{ik}$ is the degree of node $i$, then we have that 
\begin{equation*}
	\dot{u}_i = - d_i u_i + \sum_{k=1}^n A_{ik} u_k \;.
\end{equation*}
Written in matrix form, this expression translates to 
\begin{equation*}
	\dot{\mathbf{u}} = - D \mathbf{u} \,+\, A \mathbf{u} = \left(-D+A \right) \mathbf{u} = L \mathbf{u} \;.
\end{equation*}

Consider the sample graph of $n=8$ nodes in Fig.~\ref{fig:Figure5}B. If all nodes are at the same temperature, e.g., $u_i(0) = 5^\circ$C, then all pair-wise differences between adjacent nodes in Eq.~(\ref{eq:Laplacian1}) vanish, $A_{ik} \, (u_k - u_i) = 0$, and no heat flows. If the temperature of the first node suddenly increased to $u_1 = 15^\circ$C, heat will start to flow towards its adjacent neighbors $i=2$ and $i=3$, increasing $u_2$ and $u_3$, and decreasing $u_1$. This will extend to the rest of the network through the neighbors of $i=2$ and $i=3$. Temperature $u_1$ will decrease and all other $u_i$ will increase until all nodes reach $6.25^\circ$C, Fig.~\ref{fig:Figure5}C, which is the average temperature of the initial conditions $\mathbf{x}_0^T = \left(15, 5, 5, \ldots, 5 \right)$.

\section*{Appendix 4: Relation between time-discrete and time-continuous random walks}	\label{sec:Appendix4}

In Section~V we reviewed several different approaches proposed in the literature that employ the random walk to explore networks and derive centrality measures or identify communities. Some works employ the simple random walk $\mathbf{x}_{t+1} = T \, \mathbf{x}$ where $T$ is the transition probability matrix. Other proposals are based on the so-called \emph{time-continuous random walk} by equation $\dot{\mathbf{p}} = [- \mathrm{I} + T] \, \mathbf{p} = \mathcal{L} \, \mathbf{p}$, where $\mathcal{L} \equiv D^{-1}L = -I + T$ is the normalized Laplacian matrix. We showed that this time-continuous equation is the same as the heat equation $\dot{\mathbf{x}} = L \mathbf{x}$ only that the entries of $\mathcal{L}$ are column-wise normalized such that variables $p_i(t)$ represent probabilities bounded between 0 and 1.

\begin{figure}
	\centering
	\includegraphics[width=1.0\columnwidth,clip=]{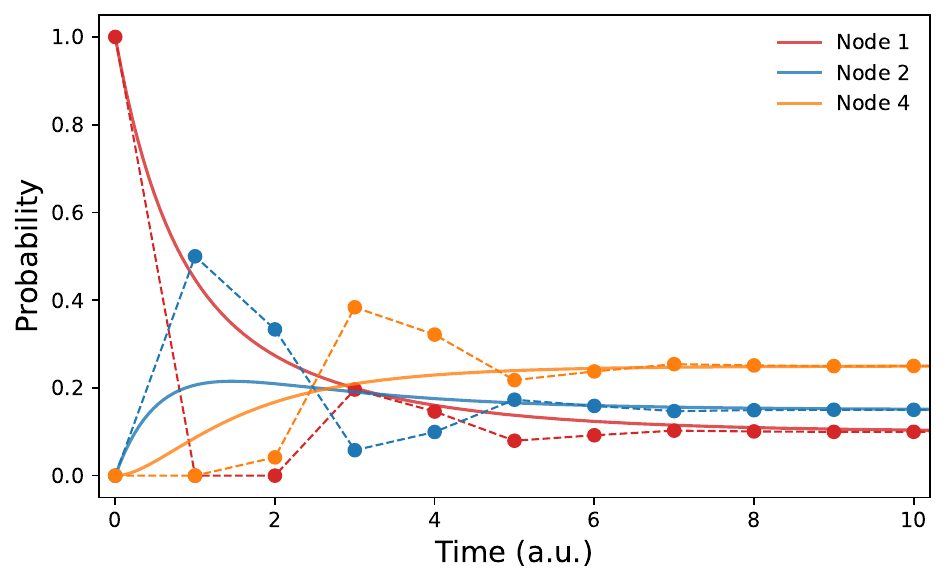}
 	\caption{		\label{fig:Figure6}
	{\bf Relation between time-discrete and time-continuous random walks.}
	Temporal evolution of the probability of a random walking agent to be found at nodes $i=1, 2$ and $4$, for a single agent initially seeded on $i=1$ and navigating through the sample graph of $n=8$ nodes of Fig.~2B. Solid lines (---) represent the solutions for the time-continuous equation and dashed lines ($--$) the solutions for the time-discrete version. As evidenced, time-continuous and time-discrete solutions for each node converge at the steady-state ($t \to \infty$) but follow differentiated trajectories during the transient times. 
	} 
\end{figure}

It is informative to clarify that network metrics derived from time-discrete or time-continuous versions are equivalent, only when the metrics capture the steady-state horizon. Solutions $\mathbf{x}_t$ and $\mathbf{p}(t)$ converge in the $t \to \infty$ time limit but differ during the transient times. In order to illustrate this, we solved both equations for one random walker initially seeded at node $i=1$ (initial conditions $\mathbf{x}_0 = \mathbf{p}_0 = [1,0,0, \ldots,0]$) in the sample graph of $n=8$ illustrated in Fig.~2B. The temporal solutions (probability of a node to be visited by the agent) are shown in Fig.~\ref{fig:Figure6}. As it is observed, the time-discrete (dashed lines) and time-continuous (solid lines) solutions for a node begin from same initial conditions (at $t=0$) and converge again at the later times, but the two solutions follow separate trajectories in between. This observation illustrates the importance of choosing a \emph{right} canonical model that respects fundamental constraints and conditions of a real system. In this case, even if the models are equivalent, considering the system to be time-discrete of time-continuous leads to different propagation dynamics in the network, and therefore to different views of the influences between nodes.

\end{document}